\documentclass[twocolumn]{aastex631}
\usepackage{CLASSY_Preamble}

\begin{document}
\submitjournal{AASJournal ApJ}
\shortauthors{Arellano-C\'ordova et al.}
\shorttitle{CLASSY~V}
\title{CLASSY V: The impact of aperture effects on the inferred nebular properties of local star-forming galaxies}

\author[0000-0002-2644-3518]{Karla Z. Arellano-C\'{o}rdova}
\affiliation{Department of Astronomy, The University of Texas at Austin, 2515 Speedway, Stop C1400, Austin, TX 78712, USA}

\author[0000-0003-2589-762X]{Matilde Mingozzi}
\affiliation{Space Telescope Science Institute, 3700 San Martin Drive, Baltimore, MD 21218, USA}

\author[0000-0002-4153-053X]{Danielle A. Berg}
\affiliation{Department of Astronomy, The University of Texas at Austin, 2515 Speedway, Stop C1400, Austin, TX 78712, USA}

\author[0000-0003-4372-2006]{Bethan L. James}
\affiliation{AURA for ESA, Space Telescope Science Institute, 3700 San Martin Drive, Baltimore, MD 21218, USA}

\author[0000-0002-0361-8223]{Noah S. J. Rogers}
\affiliation{Minnesota Institute for Astrophysics, University of Minnesota, 116 Church St. SE, Minneapolis, MN 55455;}

\author[0000-0003-4137-882X]{Alessandra Aloisi}
\affiliation{Space Telescope Science Institute, 3700 San Martin Drive, Baltimore, MD 21218, USA}

\author[0000-0001-5758-1000]{Ricardo O. Amor\'{i}n}
\affiliation{Instituto de Investigaci\'{o}n Multidisciplinar en Ciencia y Tecnolog\'{i}a, Universidad de La Serena, Raul Bitr\'{a}n 1305, La Serena 2204000, Chile}
\affiliation{Departamento de Astronom\'{i}a, Universidad de La Serena, Av. Juan Cisternas 1200 Norte, La Serena 1720236, Chile}

\author[0000-0003-4359-8797]{Jarle Brinchmann}
\affiliation{Instituto de Astrof\'{i]}sica e Ci\^{e}ncias do Espa\c{c}o, Universidade do Porto, CAUP, Rua das Estrelas, PT4150-762 Porto, Portugal}

\author[0000-0003-3458-2275]{St\'{e}phane Charlot}
\affiliation{Sorbonne Universit\'{e}, CNRS, UMR7095, Institut d'Astrophysique de Paris, F-75014, Paris, France}

\author[0000-0002-0302-2577]{John Chisholm}
\affiliation{Department of Astronomy, The University of Texas at Austin, 2515 Speedway, Stop C1400, Austin, TX 78712, USA}

\author[0000-0003-1127-7497]{Timothy Heckman}
\affiliation{Center for Astrophysical Sciences, Department of Physics \& Astronomy, Johns Hopkins University, Baltimore, MD 21218, USA}

\author[0000-0003-3334-4267]{Stefany Fabian Dub\'on}
\affiliation{Department of Physics, Bryn Mawr College, Bryn Mawr, PA 19010}

\author[0000-0001-8587-218X]{Matthew Hayes}
\affiliation{Stockholm University, Department of Astronomy and Oskar Klein Centre for Cosmoparticle Physics, AlbaNova University Centre, SE-10691, Stockholm, Sweden}

\author[0000-0003-4857-8699]{Svea Hernandez}
\affiliation{AURA for ESA, Space Telescope Science Institute, 3700 San Martin Drive, Baltimore, MD 21218, USA}

\author[0000-0001-5860-3419]{Tucker Jones}
\affiliation{University of California, Davis}

\author[0000-0002-5320-2568]{Nimisha Kumari}
\affiliation{AURA for ESA, Space Telescope Science Institute, 3700 San Martin Drive, Baltimore, MD 21218, USA}

\author[0000-0003-2685-4488]{Claus Leitherer}
\affiliation{Space Telescope Science Institute, 3700 San Martin Drive, Baltimore, MD 21218, USA}

\author[0000-0001-9189-7818]{Crystal L. Martin}
\affiliation{Department of Physics, University of California, Santa Barbara, Santa Barbara, CA 93106, USA}

\author[0000-0003-2804-0648]{Themiya Nanayakkara}
\affiliation{Swinburne University of Technology, Melbourne, Victoria, AU}

\author[0000-0003-1435-3053]{Richard W. Pogge}
\affiliation{Department of Astronomy, The Ohio State University, 140 W 18th Avenue, Columbus, OH 43210, USA}

\author[0000-0003-4792-9119]{Ryan Sanders}
\affiliation{University of California, Davis}

\author[0000-0002-9132-6561]{Peter Senchyna}
\affiliation{Carnegie Observatories, 813 Santa Barbara Street, Pasadena, CA 91101, USA}

\author[0000-0003-0605-8732]{Evan D. Skillman}
\affiliation{Minnesota Institute for Astrophysics, University of Minnesota, 116 Church Street SE, Minneapolis, MN 55455, USA}

\author[0000-0002-6172-733X]{Dan P. Stark}
\affiliation{Steward Observatory, The University of Arizona, 933 N Cherry Ave, Tucson, AZ, 85721, USA}

\author[0000-0001-8289-3428]{Aida Wofford}
\affiliation{Instituto de Astronom\'{i}a, Universidad Nacional Aut\'{o}noma de M\'{e}xico, Unidad Acad\'{e}mica en Ensenada, Km 103 Carr. Tijuana-Ensenada, Ensenada 22860, M\'{e}xico}

\author[0000-0002-9217-7051]{Xinfeng Xu}
\affiliation{Center for Astrophysical Sciences, Department of Physics \& Astronomy, Johns Hopkins University, Baltimore, MD 21218, USA}
\correspondingauthor{Karla Z. Arellano-C\'ordova} 
\email{kzarellano@utexas.edu}


\begin{abstract}
Strong nebular emission lines are an important diagnostic tool for tracing the evolution of star-forming galaxies across cosmic time. However, different observational setups can affect these lines, and the derivation of the physical nebular properties. We analyze 12 local star-forming galaxies from the COS Legacy Spectroscopy SurveY (CLASSY) to assess the impact of using different aperture combinations on the determination of the physical conditions and gas-phase metallicity. We compare optical spectra observed with the SDSS aperture, which has a 3$\arcsec$ diameter similar to COS, to IFU and longslit spectra, including new LBT/MODS observations of five CLASSY galaxies. We calculate the reddening, electron densities and temperatures, metallicities, star formation rates, and equivalent widths (EWs).  We find that measurements of the electron densities and temperatures, and metallicity remained roughly constant with aperture size, indicating that the gas conditions are relatively uniform for this sample. However, using the IFU observations of 3 galaxies, we find that the E(B-V) values derived from the Balmer ratios decrease ( by up to 53\%) with increasing aperture size. The values change most significantly in the center of the galaxies, and level out near the COS aperture diameter of 2\farcs5. We examine the relative contributions from the gas and stars using the H$\alpha$ and [\ion{O}{3}] \W5007  EWs as a function of aperture light fraction, but find little to no variations within a given galaxy. These results imply that the optical spectra provide nebular properties appropriate for the FUV CLASSY spectra, even when narrow 1\farcs0 long-slit observations are used.
\end{abstract} 

\section{Introduction}\label{sec:1}

A key goal of galaxy evolution is to understand the main processes shaping
the stellar and nebular gas content of galaxies across cosmic time. 
In the coming years, the new generation of space and ground-based telescopes, 
such as the James Web Space Telescope (\textit{JWST}) and the Extremely Large 
Telescopes (\textit{ELTs}) will provide our the first 
probes of the physical conditions in the first galaxies. 
In this context, the study of nearby, chemically-young, high-ionization dwarf 
galaxies provide a more detailed view of conditions similar to these high-$z$ 
systems \citep[e.g.,][]{senchyna17, senchyna19, berg21a}, 
and, thus, help to constrain and interpret their physical properties.
In particular, optical emission spectra of local, star-forming, dwarf galaxies 
provide a rich source of information about the physical properties of their 
interstellar medium (ISM), such as dust content, electron density and temperature 
structure, gas-phase metallicity, and the ionization state of the gas
\citep[e.g.,][and many more]{lequeux79, campbell86, skillman89, pagel92, vilchez95, izotov99, kunth00, berg12, berg16, sanchez-almeida16, guseva17, james17, berg19, mcquinn20}.

Recently, \citet[][hereafter, \PI]{berg22} presented the COS Legacy Archive Spectroscopic SurveY 
(CLASSY) treasury, obtained with Cosmic Origins Spectrograph (COS) on the Hubble Space Telescope (HST). 
CLASSY comprises a sample of 45 local ($0.002<$ $z$ $<0.182$) star-forming galaxies with 
high-resolution ($R \sim 15,000$) far-ultraviolet (FUV) COS spectra and moderate-resolution
($R \sim 2,000$) optical spectra.
Most of the optical observations are archival spectra from the Sloan Digital Sky Survey 
Data Release 12 (SDSS) \citep{eisenstein11}, but is complimented by long-slit and integral 
field unit spectroscopy (IFU) spectroscopy for some CLASSY galaxies 
(e.g., ESO 096.B-0923; PI Östlin, ESO 0103.B-0531; PI Erb and \citet{senchyna19, sanders22}).
The CLASSY sample covers a broad range of physical properties such as 
reddening ($0.02<$ $E(B-V)$ $<0.67$), 
nebular density, ($10<n_e$ (cm$^{-3}) <1120$), 
gas-phase metallicity ($7.0<$ 12+log(O/H) $<8.7$), 
ionization parameter ($0.5<$ [\ion{O}{3}] \W5007/[\ion{O}{2}] \W3727 $<38.0$), 
stellar mass ($6.2<$ log $M_\star$ ($M_\odot$) $<10.1$), and 
star formation rate ($-2.0<$ log SFR ($M_\odot$ yr$^{-1}$)  $<+1.6$).
One of the main purposes of CLASSY is to create a database of local star-forming galaxies 
to study and interpret the physical properties of the stellar and gas-phase content of 
star-forming galaxies across all redshifts.   
However, in contrast to distant galaxies, a large fraction of the 
light of these very nearby objects can fall out of the aperture of the fiber or slit used, 
resulting in aperture effect issues.

While optical spectra of individual \ion{H}{2} regions in nearby galaxies have provided the 
foundation for diagnosing the physical conditions in nebular gas, it is not yet clear 
whether the same diagnostics are appropriate for the 
integrated-light spectra observed from distant galaxies \citep[e.g.,][]{kobulnicky99,moustakas06}.
Several studies have focused their attention on the combined impact of aperture size with different
galaxy morphologies on the inferred physical properties of local and high-redshift galaxies  
\citep[e.g,][]{zaritsky95, perezgonzalez03, Gomez03, hopkins03, brinchmann04, kewley05}. 
For example, \citet{kewley05} presented an analysis of the aperture effects on the computation of SFR, 
reddening ($E(B-V)$), and gas-phase metallicity for different Hubble-type galaxies. 
These authors concluded that apertures capturing $<$ 20\% of the total galaxy's light show 
significant differences in the determination of the global galaxy properties.
Therefore, aperture effects can strongly bias comparisons among different surveys \citet{kewley05}. 
As such, aperture corrections are fundamental for small apertures to avoid biases \citep{hopkins03, nakamura04, brinchmann04}. 

Recently, \citet{Mannucci21} analyzed the impact of the different aperture sizes 
(e.g., representative apertures for long-slit and IFU spectroscopy) from pc to kpc scales
in nearby galaxies.
These authors found significant differences in the flux ratios involving low-ionization emission lines 
(i.e., [\ion{S}{2}] \W\W6717,6731/H$\beta$) mainly due to the internal structure of \ion{H}{2} regions. 
While we cannot resolve individual \ion{H}{2} regions in most dwarf star-forming galaxies, 
aperture effects may still play an important role when different instruments/apertures 
sample different physical sizes of the same galaxy.
Therefore, aperture effects can lead to significant biases that affect the computation of  
physical properties in both local star-forming galaxies and high redshift galaxies~\citep{kewley05, 
perezgonzalez03, Gomez03, brinchmann04}.
For this reason, it is important to understand and characterize the effects of aperture size 
on the interpretation of galaxy properties derived from rest-frame optical spectra. 

In this paper, we present an analysis of the 12 CLASSY galaxies for which 
multiple optical spectra exist with different apertures.
The variations in spatial scales probed by these different spectra provide an ideal laboratory 
to assess the variations of physical properties.
 Therefore, this sample provides an excellent opportunity to assess whether differences in aperture 
size or instrument produce significantly different, optically-derived properties. 
In the local Universe, optical emission lines are the main source of information to obtain reliable 
measurements of the gas-phase metallicity, the ionization state of the gas, star formation rate, and 
other valuable nebular properties.
However, the study of FUV spectra is crucial to fully characterize the young stellar 
populations and their impact of the nebular properties. 
Therefore, a deeper understanding of galaxy evolution requires a joint analysis of 
the FUV$+$optical CLASSY spectra.
Given the different instruments and apertures used for the CLASSY FUV$+$optical spectra, 
it is important to evaluate the potential bias introduced by such observations. 
With our sub-sample of 12 CLASSY star-forming galaxies, we investigate the effect of 
aperture sizes on optical spectra derived nebular properties. 
Further, this analysis is crucial to determine whether our optically-derived
nebular properties are appropriate for the 2\farcs5-diameter aperture of the 
HST/COS spectra.

The structure of this paper is as follows. We describe our sample and the optical observation in Section~\ref{sec:sample}. In Section~\ref{sec:ebv}, we present the analysis of observations, the computation of the reddening, and aperture comparison. In Sections~\ref{sec:abundances}, we calculate physical conditions, metallicities and equivalent widths (EWs), and the ionization parameter of the gas for the multiple apertures. In Section~\ref{sub_sec:discussion}, we compare and discuss the results and finish with a summary and conclusion in Section~\ref{sec:conclusion}. 

\section{Optical Spectra}\label{sec:sample}
The ancillary optical spectra of CLASSY comprises observations of APO/SDSS \citep{eisenstein11} of 38 out of 45 objects, and multiple observations of Keck/ESI and VLT/MUSE IFU \citep{sanders22, senchyna19} for a small sub-sample of the CLASSY galaxies \citep{berg22}. 
In addition, seven CLASSY galaxies were observed using long-slit spectroscopy with the Large Binocular Telescope (LBT) \citep{hill10} to ensure the accurate measurement of the physical properties of these galaxies (see Section~\ref{sec:LBT_spectra}). In summary, the CLASSY sample is fully covered by optical spectra for different observational sets available \citep{berg22,james22} and follow-up observations.
The telescope and instruments used for the optical observations in this paper are listed in Table~\ref{tab:journal}.

In this work, we have selected the 12 star-forming galaxies from the CLASSY sample 
having optical spectra with multiple aperture sizes (1.0\arcsec\,- 3.0\arcsec\,), observational modes (long-slit and IFU model), and spectral resolutions. 
In particular, the IFU data allow us to map the gas emission using different aperture sizes, which provides a better comparison between the different observations of this sample, e.g., to analyze the results obtained using the same aperture as in HST/COS \citep{berg22, james22}. 

\begin{figure*} 
\begin{center}
    \includegraphics[width=0.80\textwidth, trim=30 0 30 0,  clip=yes]{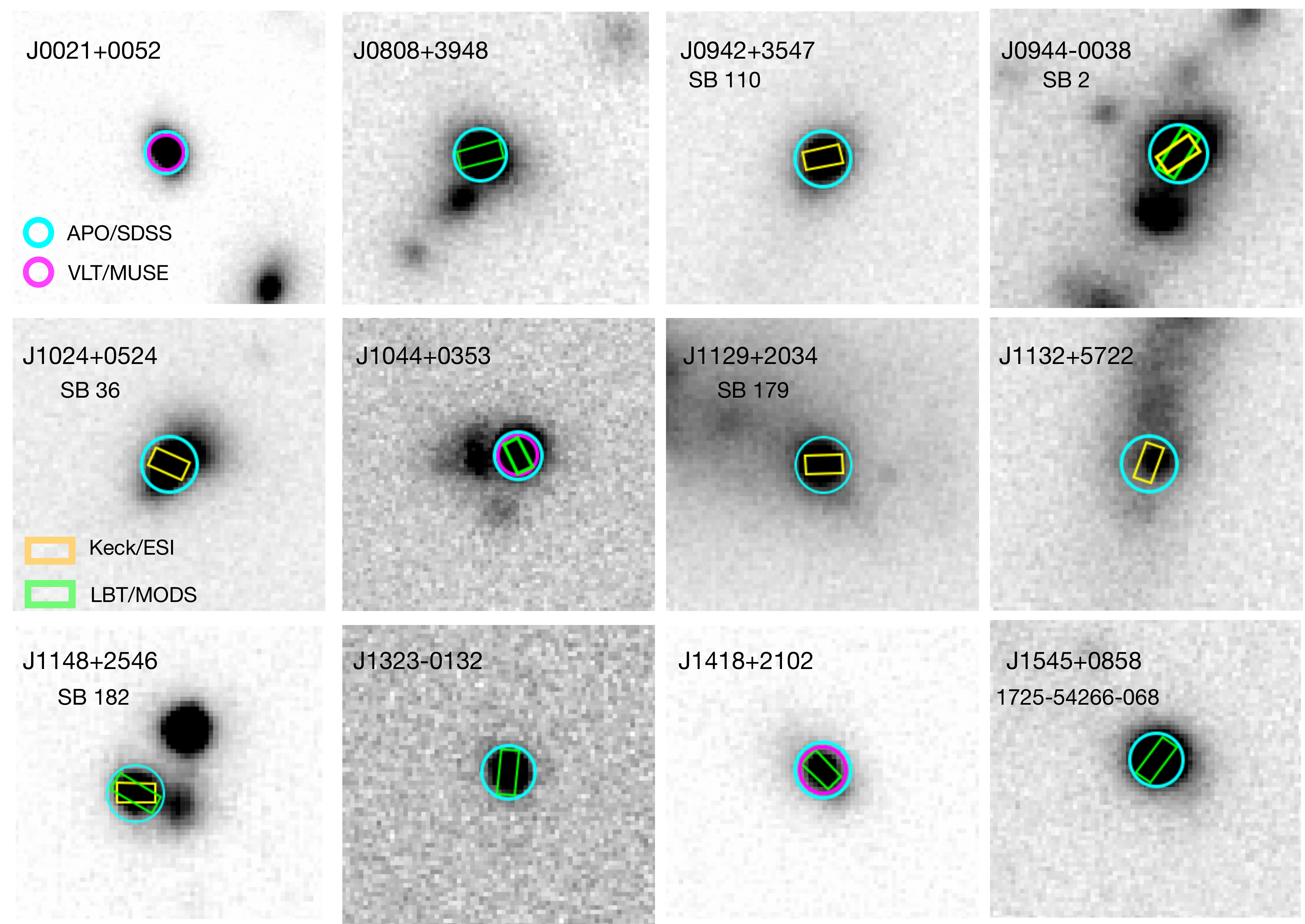}
    \caption{Field of View —optical images of Pan-STARRS in r-band for 12 star-forming galaxies of CLASSY. The 3\arcsec\ SDSS aperture used for the optical spectra is shown as a cyan circle and the 2.5\arcsec\ MUSE aperture as a magenta circle, whose diameter and pointing is the same as the HST/COS 2.5\arcsec\ aperture. We highlight for a comparison also the 1\arcsec $\times$ 2.5\arcsec\ LBT/MODS slit (green box) and 1\arcsec $\times$ 2\arcsec\ Keck/ESI slit (orange box). The position of the apertures is centred at the brightest emission knot of each galaxy using SDSS r-images. The labels show the ID of each galaxy used in this study and the alternative name for some objects. North up and East to the left. }
    \label{fig:apertures}
  \end{center}
\end{figure*}

In Figure~\ref{fig:apertures}, we show the optical images of Pan-STARRS\footnote{https://panstarrs.stsci.edu/} in the r-band for the 12 galaxies of this study. We also overlay the aperture at the observed position angle. SDSS and MUSE observations are indicated with circular apertures of 3\arcsec\, and 2.5\arcsec\, of diameter, respectively. ESI and LBT apertures have an size of 1\arcsec wide with an 1D extraction box of $\sim$ 1\arcsec\, $\times$ 2\arcsec\ and 1\arcsec\, $\times$  2.5\arcsec\, respectively. In Fig.~\ref{fig:apertures}, we show the extraction sampling for each aperture. Although most of the galaxies show extended emission, the multiple aperture are covering the bright emission region of each galaxy. 

To identify and compare the different aperture observations, we simply refer to the observations as SDSS, LBT, MUSE and ESI for the rest of the paper.

\subsection{LBT/MODS Observations}\label{sec:LBT_spectra}
\label{sec:LBT}
Here we present new optical spectra for seven CLASSY galaxies using the Multi-Object Double Spectographs (MODS, \citealt{pogge10}) mounted on the LBT \citep{hill10}.
The LBT spectra of two of these galaxies (J1044+0353 and J1418+2102) was already reported in \citet{berg21a}. 
The optical LBT/MODS spectra of the remaining five CLASSY galaxies were obtained on the UT dates of Feb 11th-12th and March 17th-18th, 2021, respectively. MODS has a large wavelength coverage from 3200 \AA\, to 10000 \AA\, with a moderate spectral resolution of $R \sim 2000$. The blue and red spectra were obtained simultaneously using the G400L (400 lines mm$^{-1}$) at $R \approx 1850$ and G670L (250 lines mm$^{-1}$) at $R \approx 2300$ gratings. The slit length was 60 arcsec with a width of 1 arcsec. To minimize flux losses due to atmospheric differential refraction \citep{filippenko82}, the slit was oriented along the parallactic angle at half the total integration at airmasses ranging between 1.05 and 1.31. Each object was observed with a total exposure time of 45 min (3$\times$900s exposures).
Fig.~\ref{fig:apertures} shows the slit position and orientation of the LBT/MODS observations for J0808+3948, J0944-0038, J1148+2546 (SB\,182), J1323-0132 and J1545+0858 (1725-54266-068). For comparison purposes, the LBT/MODS slit has been truncated (1\arcsec\ $\times$ 2.5\arcsec\,) to show the  extraction region of the observation.
The slit was centered on the highest surface brightness knots of optical emission based on SDSS r-band images, following the position of HST/COS aperture of each CLASSY galaxy in the UV \citep{berg22}.

\begin{figure*} 
\begin{center}
    \includegraphics[width=0.97\textwidth, trim=30 0 30 0,  clip=yes]{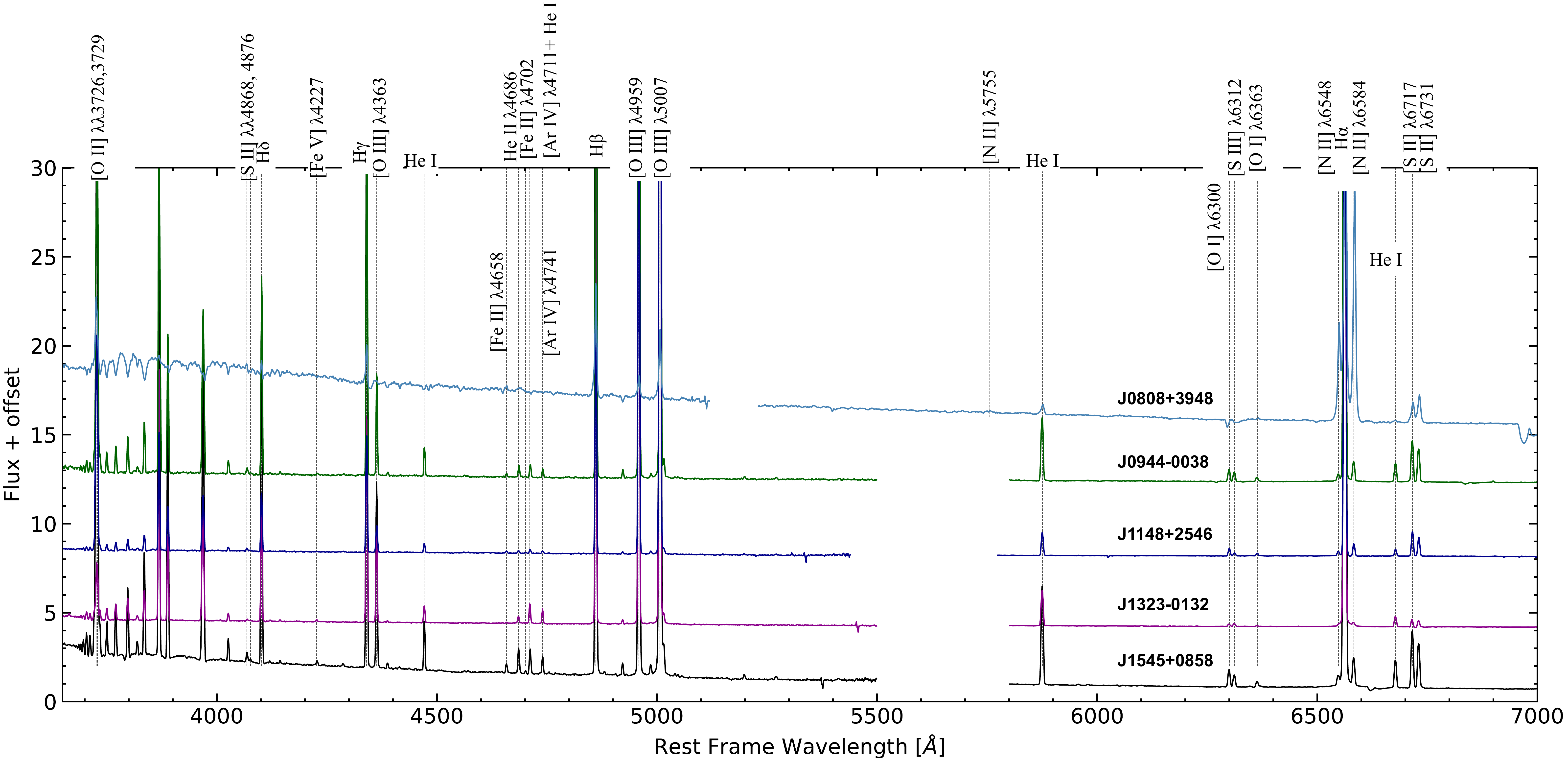}
    \caption{New optical LBT/MODS spectra of five CLASSY galaxies. The dotted lines indicate the main emission lines used in this study. F$_\lambda$ is units of $10^{-16}$ ergs$^{-1}$ s$^{-1}$ cm$^{-2}$ \AA\,$^{-1}$. We added an offset vertically for display with exception of J1545+0858. The offset corresponds to 4, 8, 12, 14 for J1323-0132, J1148+2546, J0944-0038 and J0808+3948, respectively. The emission lines for these spectra used in this study are reported in Table~\ref{tab:LBT_intensities}, and the complete version of all emission lines of MODS/LBT spectra are reported in \citet{mingozzi22}.}
    \label{fig:spectralbt}
  \end{center}
\end{figure*}

We use the CHAOS project's
data reduction pipeline \citep{berg15,croxall16, Rogers21} to reduce LBT/MODS observations. To correct for bias and flat field we apply the modsCCDRed\footnote{https://github.com/rwpogge/modsCCDRed} Python programs to the  standard stars, science objects and calibration lamps. The resulting CCD images are used in the beta version of the MODS reduction pipeline\footnote{http://www.astronomy.ohio-state.edu/MODS/Software/modsIDL/} which runs within the XIDL \footnote{https://www.ucolick.org/~xavier/IDL/} reduction package. We perform sky subtraction, wavelength calibration and flux calibration using the standard stars Feige\,34, Feige\,67 and G191-B2B. In Fig.~\ref{fig:spectralbt}, we show the one-dimensional spectra for these galaxies. We have labeled the main emission line features. Note that with the exception of J0808+3948, the rest of the galaxies shows the very high ionization emission line \ion{He}{2} \W4686 \citep{berg21a}. 

\subsection{Archival Optical Spectra}
\label{sec:arcillary}
Berg et al. (2022) provide a detailed description of the optical observations for CLASSY, while the scaling of the optical data to UV data is presented in Mingozzi et al. (2022). Here, we give a brief summary of the main characteristics of the CLASSY galaxies relevant for this study. In Table~\ref{tab:journal}, we list the optical observations of the selected galaxies. It includes the identification of the galaxy that we use in this work, the alternative names used in the literature, the coordinates, the observational ancillary, aperture size extraction of the observations and the optical size of each galaxy represented
as the total galaxy radius, ($\sim$ $R_{100}$/2) (see Sec. \ref{sec:physical_properties}).

\textit{APO/SDSS}: SDSS sample comprises 12 observations of 3\arcsec\ fiber diameter with a range in wavelength of 3800-9200 \AA\ and spectral resolution of $R \approx 1500-2500$ \citep{eisenstein11}. This wavelength range allows to measure the main emission lines necessary for the analysis of the physical properties of these galaxies. We use this sample as reference to compare with the rest of the observations since the aperture size and center are similar to those of HST/COS data (Berg et al. 2022).

\textit{VLT/MUSE}: we gathered archival VLT/MUSE IFU data for three CLASSY galaxies J0021+0052 (PI: Anderson), J1044  and J1418 (PI: Dawn Erb) (see Table~\ref{tab:journal}). These observations cover a wavelength from 4300 \AA\, to 9300 \AA\, at a spectral resolution of $R \approx 2000-3500$, and a field of view (FoV ) of of 1' $\times$1'.
For the analysis of this sample, we have extracted one dimensional spectrum using the same 2.5\arcsec\ HST/COS aperture.  
As an extra analysis, we have extracted one-dimensional spectra using different aperture sizes ranging from 1\arcsec to 4\arcsec\, of diameter in steps of 0.5\arcsec. With such spectra, we study the variations of the physical properties with respect to the aperture size. This analysis is discussed in Section~\ref{sec:muse-apertures}. 

We refer to this sample as VLT/MUSE or IFU to distinguish from long-slit spectra. Note that we have extracted the integrated flux using an aperture size of 2.5\arcsec\ resulting in one-dimensional spectra.

\textit{Keck/ESI}: This sample comprises six CLASSY galaxies with high spectral resolution using the Echellette Spectrograph and Imager (ESI) on Keck II. These spectra cover a rest frame wavelength from 3800 \AA\, to 10000 \AA\, approximately, at a spectral resolution of $R\approx4700$ at 1\arcsec\ slit width. The long-slit spectra of these galaxies were obtained from \citet{senchyna19} and \citet{sanders22}.

\begin{deluxetable*}{lccccccccc}
\setlength{\tabcolsep}{2pt}
\tablecaption{Multiple Optical Observations the 12 CLASSY star-forming galaxies analyzed in this work. \label{tab:journal}} 
\tablehead{
       &  Alt.                   &   R.A.            &         Decl.      &  $z$           &Telescope          &  Aperture        & Aperture size      & $R_{100}$/2 & Reference  \\ 
Galaxy &  Name                   &   (hh:mm:ss)      &  ( $\pm$dd:mm:ss)  &                &/instrument        &   extraction    &  (kpc)          & (\arcsec) &             \\ 
}
\startdata
1. J0021+0052   &                         &  00:21:01.03    &   $+$00:52:48.08  &   0.098     & APO/SDSS      & 3\arcsec\   circ                 &   5.43 &    1.66          & \citet{eisenstein11} \\
                &                         &                 &                   &             & VLT/MUSE      & 2.5\arcsec\ circ                 &   4.52 &                  & ESO 0104.D-0503; PI Anderson\\
2. J0808+3948   &                         &  08:08:44.28    &   $+$39:48:52.51  &   0.0912    & APO/SDSS      & 3\arcsec\   circ                 &   5.10 &    2.90          & \citet{eisenstein11}\\ 
                &                         &                 &                   &             & LBT/MODS      & 1\arcsec\ $\times$ 2.5\arcsec\   &   1.70 &                  & This Work \\
3. J0942+3547   & CG-274, SB\,110         &  09:42:52.78    &   $+$35:47:25.98  &   0.0149    &APO/SDSS       & 3\arcsec\ circ                   &   0.91 &     3.4          & \citet{eisenstein11}\\
                &                         &                 &                   &             & Keck/ESI      & 1\arcsec\ $\times$ 2\arcsec      &   0.30 &                  & \citet{sanders22}\\   
4. J0944-0038   & CGCG007-025, SB\,2      &  09:44:01.87    &   $-$00:38:32.18  &   0.005     & APO/SDSS      & 3\arcsec\ circ                   &   0.31 &   2.18           & \citet{eisenstein11}\\   
                &                         &                 &                   &             & Keck/ESI      & 1\arcsec\ $\times$ 2\arcsec\     &   0.10 &                  & \citet{sanders22}\\ 
                &                         &                 &                   &             & LBT/MODS      &  1\arcsec\ $\times$ 2.5\arcsec\  &   0.10 &                  & This work\\
5. J1024+0524   & SB\,36                  &  10:24:29.25    &   $+$05:24:51.02  &   0.033     & APO/SDSS      & 3\arcsec\ circ                   &   1.98 &    2.75          &\citet{eisenstein11}\\ 
                &                         &                 &                   &             & Keck/ESI      & 1\arcsec\ $\times$ 2\arcsec\     &   0.66 &                  & \citet{sanders22}\\  
6. J1044+0353   &                         &  10:44:57.79    &   $+$03:53:13.10  &   0.0129    & APO/SDSS      & 3\arcsec\ circ                   &   0.80 &    3.10          & \citet{eisenstein11} \\ 
                &                         &                 &                   &             & VLT/MUSE      & 2.5\arcsec\ circ                 &   0.66 &                  & ESO 0103.B-0531; PI Erb \\
                &                         &                 &                   &             & LBT/MODS      & 1\arcsec\ $\times$ 2.5\arcsec\   &   0.26 &                  & \citet{berg21a}\\
7. J1129+2034   & SB\,179                 & 11:29:14.15     &  $+$20:34:52.01   &   0.005     & APO/SDSS      & 3\arcsec\ circ                   &   0.31 &    4.20          & \citet{eisenstein11} \\
                &                         &                 &                   &             & Keck/ESI      & 1\arcsec\ $\times$ 2\arcsec\     &   0.10 &                  & \citet{sanders22}\\  
8. J1132+5722   & SBSG1129+576            & 11:32:35.35     & $+$57:22:36.39    &  0.018      & APO/SDSS      & 3\arcsec\ circ                   &   1.10 &    4.05          & \citet{eisenstein11} \\ 
                &                         &                 &                   &             & Keck/ESI      &  1\arcsec\ $\times$ 2\arcsec\    &   0.37 &                  & \citet{senchyna19}\\
9. J1148+2546   &  SB\,182                &  11:48:27.34    &  $+$25:46:11.77   &   0.045     & APO/SDSS      & 3\arcsec\ circ                   &   2.65 &    1.33          & \citet{eisenstein11}\\ 
                &                         &                 &                   &             & Keck/ESI      & 1\arcsec\ $\times$ 2\arcsec\     &   0.88 &                  & \citet{sanders22}\\
                &                         &                 &                   &             & LBT/MODS      & 1\arcsec\ $\times$ 2.5\arcsec\   &   0.88 &                  & This work\\
10. J1323-0132  &                         &  13:23:47.52    & $-$01:32:51.94    &  0.022      & APO/SDSS      & 3\arcsec\ circ                   &   1.33 &    1.66          & \citet{eisenstein11} \\ 
                &                         &                 &                   &             & LBT/MODS      & 1\arcsec\ $\times$ 2.5\arcsec\   &   0.44 &                  & This work\\
11. J1418+2102  &                         &  14:18:51.12    & $+$21:02:39.84    &  0.009      & APO/SDSS      & 3\arcsec\ circ                   &   0.55 &    2.84          & \citet{eisenstein11}\\ 
                &                         &                 &                   &             & VLT/MUSE      & 2.5\arcsec\ circ                 &   0.46 &                  & ESO 0103.B-0531; PI Erb \\
                &                         &                 &                   &             & LBT/MODS      & 1\arcsec\ $\times$ 2.5\arcsec\   &   0.18 &                  & \citet{berg21a}\\
12. J1545+0858  & 1725-54266-068          &  15:45:43.44    & $+$08:58:01.34    &   0.038     & APO/SDSS      & 3.0\arcsec\ circ                 &   2.26 &    2.54          & \citet{eisenstein11}\\ 
                &                         &                 &                   &             & LBT/MODS      & 1\arcsec\ $\times$ 2.5\arcsec\   &   0.75 &                  & This work\\
\enddata
\tablecomments{The different columns indicate: (1) galaxy ID, (2) alternative name, (3-4) coordinates of the objects (J2000), (5) redshift, (6) telescope, instrument or survey for each observation, (7) aperture extraction size, (8) physical size within an aperture size of 3\arcsec for SDSS, 2.5\arcsec\ for MUSE and 1\arcsec\  for LBT/MODS,  Keck/ESI, and IFU aperture of 2.5\arcsec and 3.0\arcsec apertures for VLT/MUSE and APO/SDSS, (9) the total galaxy radius represented as ($\sim$ $R_{100}$/2) and derived using $r$-band images of Pan-STARRS \citep{berg21a} (see also Sec.~\ref{sec:physical_properties})}, and (10) references of the sample.
\end{deluxetable*} 

\begin{deluxetable*}{lccccccccc }
\tablecaption{Physical properties of the 12 CLASSY star-forming galaxies from different apertures analyzed in this work. \label{tab:Evb} }
\tablewidth{0pt}
\tablehead{
          & Telescope         &                     &                 $E(B-V)$ &       &           $E(B-V)$          & EW(H$\alpha$) & EW([O III] $\lambda$5007)    \\
Galaxy    & /instrument        &  H$\alpha$/H$\beta$ & H$\gamma$/H$\beta$ &   H$\delta$/H$\beta$ & Adopted  & (\AA\,) & (\AA\,)}
\startdata
J0021+0052       & APO/SDSS     & $ 0.142 \pm 0.010 $   & $ 0.116 \pm 0.021 $   & $ 0.120 \pm 0.025 $     &  $ 0.135 \pm 0.009 $           &   $ 496 \pm 15 $     &  $ 400 \pm 9 $     \\
                 & VLT/MUSE     & $ 0.237 \pm 0.012 $   & $ 0.150 \pm 0.021 $   & \nodata                 &  $ 0.214 \pm 0.011 $           &   $ 520\pm 10 $      &  $ 430 \pm 9 $     \\
\\                                                                                                                                                                                      
J0808+3948       & APO/SDSS     & $ 0.331 \pm 0.039 $   & $ 0.007 \pm 0.124 $   & $ 0.214 \pm 0.144 $     &  $ 0.296 \pm 0.036 $           &   $ 69 \pm 2 $       &  $ 7 \pm 1 $       \\
                 & LBT/MODS     & $ 0.215 \pm 0.026 $   & $ 0.226 \pm 0.04 $    & $ 0.062 \pm 0.062 $     &  $ 0.202 \pm 0.020 $           &   $ 77 \pm 2 $       &  $ 7 \pm 1 $       \\
\\                                                                                                                                                                                    
J0942+3547      & APO/SDSS      & $ 0.021 \pm 0.006 $   & $ -0.015 \pm 0.014$    & $ -0.030 \pm 0.017 $   &  $ 0.021 \pm 0.006 $           &   $ 490 \pm 24 $     &  $ 461 \pm 9 $     \\
                & Keck/ESI      & $ 0.169 \pm 0.003 $   & $ 0.063 \pm 0.008 $    & $ 0.007 \pm 0.007  $   &  $ 0.134 \pm 0.003 $           &   $ 453 \pm 19 $     &  $ 424 \pm 10 $     \\
\\                                                                                                                                                                                    
J0944-0038      & APO/SDSS      & \nodata               & $ 0.201 \pm 0.014 $   & $ 0.185 \pm 0.014 $     &  $ 0.193 \pm 0.009 $           &    \nodata           &  $ 1464 \pm 68 $   \\
                & Keck/ESI      & $ 0.169 \pm 0.003 $   & $ 0.063 \pm 0.008 $   & $ 0.007 \pm 0.007 $     &  $ 0.134 \pm 0.003 $           &   $ 1788 \pm 46 $    &  $ 1798 \pm 16 $   \\
                & LBT/MODS      & $ 0.164 \pm 0.012 $   & $ 0.172 \pm 0.022 $   & $ 0.180 \pm 0.017 $     &  $ 0.170 \pm 0.009 $           &   $ 1576 \pm 31 $    &  $ 2121 \pm 21 $   \\
\\                                                                                                                                                                                     
J1024+0524       & APO/SDSS     & $ 0.001 \pm 0.007 $   & $ 0.097 \pm 0.014 $   & $ 0.064 \pm 0.017 $     &  $ 0.025 \pm 0.006 $           &   $ 523 \pm 21 $     &  $ 513 \pm 15 $     \\
                 & Keck/ESI     & $ 0.071 \pm 0.009 $   & \nodata               & $ 0.016 \pm 0.016 $     &  $ 0.055 \pm 0.008 $           &   $ 503 \pm 12 $     &  $ 515 \pm 14 $     \\
\\                                                                                                                                                                                     
J1044 +0353     & APO/SDSS      & $ 0.118 \pm 0.011 $   & $ 0.057 \pm 0.017 $   & $ 0.003 \pm 0.015 $     &  $ 0.072 \pm 0.008 $           &   $ 1620 \pm 84 $    &  $ 1212 \pm 41 $   \\
                & VLT/MUSE      & $ 0.088 \pm 0.007 $   & \nodata               &  \nodata                &  $ 0.088 \pm 0.007 $           &   $ 1664 \pm 20 $    &  $ 1509\pm 24 $   \\
                & LBT/MODS      & $ 0.086 \pm 0.009 $   & $ 0.231 \pm 0.018 $   & $ 0.231 \pm 0.015 $     &  $ 0.14 \pm 0.007 $            &   $ 1594 \pm 56 $    &  $ 1568 \pm 40 $   \\
\\                                                                                                                                                                                   
J1129+2034      & APO/SDSS      & $ 0.220 \pm 0.010 $   & $ 0.126 \pm 0.015 $   & \nodata                 &  $ 0.191 \pm 0.009 $           &   $1220\pm60$        &  $ 935 \pm 42 $   \\
                & Keck/ESI      & \nodata               & $ 0.081 \pm 0.018 $   & $ 0.175 \pm 0.020 $     &  $ 0.125 \pm 0.013 $           &   \nodata            &  $ 1440 \pm 14 $   \\
\\                                                                                                                                                                                     
J1132+5722      & APO/SDSS     & $ 0.035 \pm 0.009 $   & $ 0.07 \pm 0.024 $     & $ 0.004 \pm 0.040 $     &  $ 0.038 \pm 0.008 $           &   $ 400 \pm 23 $     &  $ 175 \pm 5 $     \\
                & Keck/ESI     & $ 0.051 \pm 0.010 $   & $ 0.199 \pm 0.064 $    & $ 0.015 \pm 0.071 $     &  $ 0.054 \pm 0.010 $           &   $ 587 \pm 56 $     &  $ 283 \pm 28 $     \\
\\                                                                                                           
J1148+2546     & APO/SDSS     & $ 0.113 \pm 0.007 $    & $ 0.147 \pm 0.013 $   & $ 0.114 \pm 0.016 $     &  $ 0.119 \pm 0.006 $            &   $ 879 \pm 26 $    &  $ 956 \pm 11$    \\
               & Keck/ESI     & $ 0.071 \pm 0.004 $    & \nodata               & $ -0.054 \pm 0.008 $    &  $ 0.071 \pm 0.004 $            &   $ 796 \pm 16 $    &  $ 973 \pm  25$     \\
               & LBT/MODS     & $  0.060 \pm 0.012 $   & $ 0.180 \pm 0.019 $   &   \nodata             &  $ 0.093 \pm 0.010 $            &   $ 1051 \pm 30$    &  $ 1510 \pm 67 $   \\
\\                                                                                                                                                                                   
J1323-0132      & APO/SDSS     & $ 0.118 \pm 0.009 $   & $ 0.202 \pm 0.017 $   & $ 0.239 \pm 0.022 $      &  $ 0.147 \pm 0.007 $            &   $ 1561 \pm 33 $   &  $ 2080 \pm 83 $   \\
                & LBT/MODS     & $ 0.109 \pm 0.010 $   & $ 0.192 \pm 0.020 $   & $ 0.204 \pm 0.015 $      &  $ 0.146 \pm 0.008 $            &   $ 1429 \pm 30 $   &  $ 2281\pm 93 $   \\
\\                                                                                                                                                                                    
J1418+2102      & APO/SDSS     & $ 0.112 \pm 0.005 $   & $ 0.154 \pm 0.017 $   & \nodata                  &  $ 0.116 \pm 0.005 $            &   $ 1237\pm 85 $   &  $ 1053 \pm 24 $    \\
                & VLT/MUSE     & $ 0.091 \pm 0.006 $   & \nodata               & \nodata                  &  $ 0.091 \pm 0.006 $            &   $ 1329\pm 22 $   &  $ 1367 \pm 10 $   \\
                & LBT/MODS     & $ 0.045 \pm 0.011 $   & $ 0.192 \pm 0.023 $   & $ 0.215 \pm 0.018 $      &  $ 0.105 \pm 0.009 $            &   $ 1121\pm 48 $   &  $ 1342 \pm 48 $   \\
\\                                                                                                                                                                                   
J1545+0858      & APO/SDSS     & $ 0.154 \pm 0.005 $   & $ 0.210 \pm 0.015 $   & $ 0.133 \pm 0.015 $      &  $ 0.156 \pm 0.004 $            &   $ 1140 \pm 33 $  &  $ 1151 \pm 38$    \\
                & LBT/MODS     & $ 0.024 \pm 0.009 $   & $ 0.160 \pm 0.016 $   & $ 0.181 \pm 0.013 $      &  $ 0.088 \pm 0.007 $            &   $ 1132 \pm 17 $  &  $ 1394 \pm 32$   \\
\enddata
\tablecomments{The different columns indicate: (1) galaxy ID, (2) instrument/telescope of the observations, (3)-(5) the $E(B-V)$ values derived using Balmer line ratios H$\alpha$/H$\beta$, H$\gamma$/H$\beta$ and H$\delta$/H$\beta$, respectively, (6) the adopted extinction value, $E(B-V)$, calculated using the weighted mean, (7) the star formation rate derived using H$\alpha$ luminosity and (8)-(9) the equivalent width for H$\alpha$ and [\ion{O}{3}]~\W5007, respectively.}
\end{deluxetable*}


\section{Emission line Measurements}\label{sec:ebv}
\subsection{Emission Line Fluxes}
To prepare the spectra for emission line measurements, the optical spectra were corrected by Galactic extinction using the \citet{green15} extinction maps included in the {\sc Python dustmaps} package \citep{green18}, and the \citet{cardelli89} reddening law. Next, since the Balmer lines are affected by stellar absorption, we model the stellar continuum using \texttt{Starlight}\footnote{{\url{www.starlight.ufsc.br}}} spectral synthesis code \citep{cidfernandes2005} and subtract it. To do so, we follow the successful continuum-subtraction method of the CHAOS survey \citep{berg15} and assume the stellar population models of \cite{bruzual03} with an initial mass function (IMF) of \citet{chabrier03}. We use a set of simple stellar populations that span 25 ages (1 Myr--18 Gyr) and six metallicities ($0.05\,Z_\odot<$ Z $< 2.5\,Z_\odot$). 

Using the continuum-subtracted spectra, we fit the emission lines with Gaussian profiles using the \texttt{Python} package \texttt{LMFIT}\footnote{Non-Linear Least-Squares Minimization and Curve-Fitting: \url{https://github.com/lmfit/lmfit-py}}. 
Specifically, we simultaneously fit sections of nearby lines ($< 200$ \AA),
while constraining the offset from line centers and the line width.
This method also allows us to simultaneously fit weak and blended lines with a
higher degree of accuracy. 
 Since the [\ion{O}{2}]~\W\W3726, 3729 lines are blended in LBT and SDSS observations, we have fitted two Gaussian profiles to constrain the flux measurements in such lines.
The errors in the flux were calculated using the expression reported in \citet{berg13} and \citet{Rogers21}.
Note that for the LBT/MODS spectra of J1044+0353 and J1418+2102,
we have adopted the line fluxes reported in \citet{berg21a}.
To ensure significant emission line detections, we only use lines with 
S/N$>3$. 

The auroral emission lines used to measure temperatures are intrinsically faint and 
so require careful consideration.
Therefore, we inspected the 
[\ion{O}{3}] \W4363, 
[\ion{N}{2}] \W5755, 
[\ion{S}{3}] \W6312, and
[\ion{O}{2}] \W\W7320,7330 auroral lines fits
to ensure a well-constrained flux measurement.
As a secondary check, we measured these faint $T_e$-sensitive lines by hand using the 
integration tool in {\texttt {\sc iraf}}\footnote{IRAF is distributed by the National 
Optical Astronomy Observatory, which is operated by the Association of Universities for 
Research in Astronomy, Inc., under cooperative agreement with the National Science Foundation.} Both the \texttt{\sc python} and {\texttt {\sc iraf}} methods produced consistent flux values.
Additionally, for high-metallicity star-forming galaxies (12+log(O/H) $>$ 8.4) 
the [\ion{O}{3}] \W4363 emission line can be contaminated by 
[\ion{Fe}{2}]\W4359 \citep[][ 12$+$log(O/H)$>8.4$]{curti+17}. 
Recently, \citet{arellano-cordova2020b} analyzed the effect of using [\ion{O}{3}] \W4363 
blended with [\ion{Fe}{2}] in the computation of temperature and metallicity in a sample 
of Milky Way and Magellanic Cloud \ion{H}{2} regions. 
These authors concluded that the use of a contaminated [\ion{O}{3}] \W4363 line can lead 
differences in metallicity up to 0.08 dex. 
However, most of objects in the present sample are metal-poor galaxies, but, we have taken 
into account the possible contamination of [\ion{O}{3}] \W4363 by [\ion{Fe}{2}]\W4359 in 
each spectrum of the sample as a precaution.
 
\begin{figure*} 
\begin{center}
    \includegraphics[width=0.90\textwidth, trim=30 0 30 0,  clip=yes]{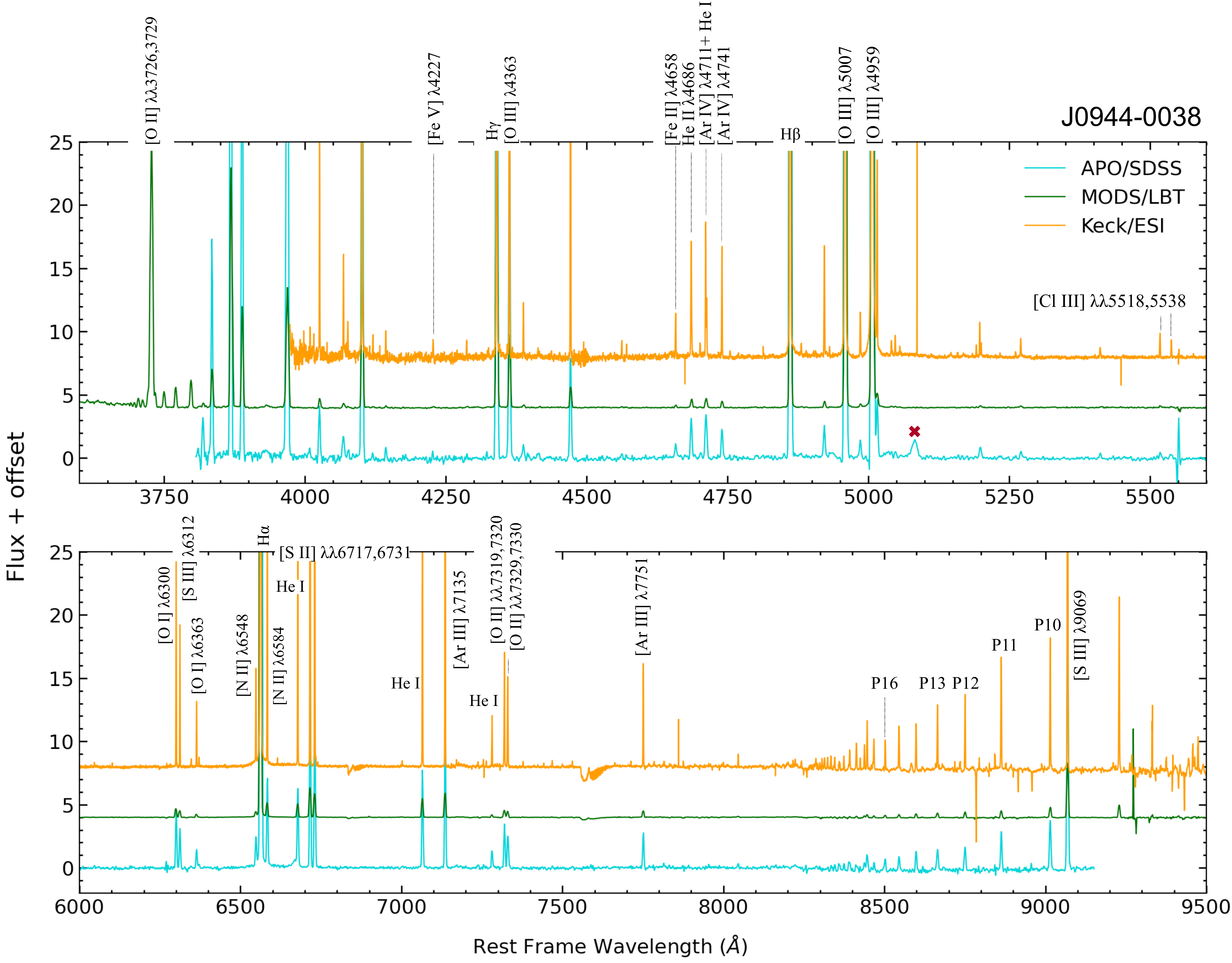}
    \caption{Comparison of the optical spectra of J0944 (or SB\,2). 
    These spectra were observed using different instruments with different aperture sizes 
    (APD/SDSS: 3\arcsec\ diameter, LBT/MODS: 1\arcsec\ slit, and Keck/ESI: 1\arcsec\ slit). 
    For a better visual comparison, the flux scale is arbitrarily offset between the spectra.
    The labels indicate the emission lines used to calculate the physical properties of star-forming galaxies. 
    In the SDSS spectrum, we identify an unknown emission artefact (red x), 
    but note that it does not affect the flux measurements of the emission lines around.
}
    \label{fig:spectra_apertures}
  \end{center}
\end{figure*}

In Fig.~\ref{fig:spectra_apertures}, we show the resulting spectrum for SDSS, LBT and ESI of J0944-0038 (or SB\,2). We have subtracted the continuum from the underlying stellar population, and we have added a small offset for a better comparison of the different spectra.

The different observations for J0944-0038 show the main emission lines used to compute the physical conditions of the nebular gas, namely the electron density and temperature and metallicity. 
These spectra contain significant \ion{He}{2} \W4686, [\ion{Ar}{4}] \W\W4711,4740, and 
[\ion{Fe}{5}] \W4227 emission, indicative of a very-high-ionization gas in this galaxy 
(see dashed lines in Fig.~\ref{fig:spectra_apertures}). 
Such very-high-ionization conditions of the gas were reported by \cite{berg21a} for 
J1044+0353 and J1418+2102. 
Additionally, 11 galaxies in our sample have \ion{He}{2} and [\ion{Ar}{4}] line detections,
but no [\ion{Fe}{5}] \W4227 detections. 
Note that for J0944-0038 (see Fig.~\ref{fig:spectra_apertures}), 
there is an odd artefact at $\approx$5100 \AA\ in the SDDS spectrum 
that is not detected in the LBT or ESI spectra. 
This feature is likely an artefact of the SDSS reduction process, 
but it does not affect our emission-line flux measurements.

\subsection{Reddening Corrections}
Before calculating nebular properties, emission lines must first be 
corrected for reddening due to dust with the galaxy.
To do so, we calculate the color-excess, $E(B-V)$, 
for three Balmer decrements (H$\alpha$/H$\beta$, H$\gamma$/H$\beta$, 
and H$\delta$/H$\beta$) using the expression:
\begin{equation}
\label{eq:ebv}
    E(B-V) = 
        \frac{-2.5}{\kappa(I_{\lambda}) - \kappa(I_{{\rm H}\beta})}\times 
        {\rm log_{10}}\left[\frac{(I_{\lambda}/I_{{\rm H}\beta})_{\rm theo.}}
                                 {(I_{\lambda}/I_{{\rm H}\beta})_{\rm obs.}} \right],
\end{equation}
where $\kappa$(\W) is the value of the attenuation curve at the corresponding 
wavelength, and $(I_{\lambda}/I_{{\rm H}\beta})_{\rm theo.}$ and 
$(I_{\lambda}/I_{{\rm H}\beta})_{\rm obs.}$ are the theoretical and observed 
Balmer ratios, respectively. 
We adopt the reddening law of \citet{cardelli89} to calculate $\kappa$(\W).

To calculate the color excesses, we use three Balmer ratios, H$\alpha$/H$\beta$, 
H$\gamma$/H$\beta$, and H$\delta$/H$\beta$, for each galaxy, with a few exceptions 
(see Sec.~\ref{sec:sample}):
\begin{itemize}[leftmargin=*,align=left]
\itemsep0em
\item For the MUSE spectra of J1044+0353 and J1418+2102, the limited spectral coverage 
limits the $E(B-V)$ calculations to only the H$\alpha$/H$\beta$ ratio.
\item For the SDSS spectra of J0944-0038, the H$\alpha$ emission line shows an asymmetric 
odd extension previously reported by \citet{senchyna17}, which is not visible in MUSE and ESI spectra.
\item For the ESI spectra of J1129+2034, the H$\alpha$ emission line is saturated and so cannot be used.
\item For the ESI spectra of J1148+2546, the H$\gamma$ emission line is affected by a detector artifact and so cannot be used.
\item For the ESI spectra of J1024+0524, both the H$\gamma$ and [\ion{O}{3}] \W4363 emission lines are affected by a detector artifact and so cannot be used.
\item For the LBT spectra of J1148+2546, the H$\delta$ is affected by an artifact and so cannot be used.
\end{itemize}

We calculate $E(B-V)$ values following the approach of \citep{berg22},
which iteratively calculates the electron temperature and density and $E(B-V)$ values.
We calculate the initial electron temperature using $T_{\rm e}$[\ion{O}{3}] [\ion{O}{3}]~(\W\W4959,5007)/\W4363 when measured, and other available temperature diagnostics when not (e.g., $T_{\rm e}$[\ion{S}{3}] for MUSE).
Using this temperature, and assuming the [\ion{S}{2}] densities reported in \citet{berg22} and Case B recombination, we calculated the theoretical ratios of H$\alpha$, H$\beta$, H$\gamma$ and H$\delta$ relative to $H\beta$. The iterative process stops when the difference in $T_{\rm e}$ $\leq 20$ K. Note that in each iteration, the line intensities used to calculate temperature and density were corrected for reddening. 
For comparison, we have also calculated the $E(B-V)$ values for the SDSS sample assuming an unique value of $T_{\rm e}$ = 10000 K and $n_{\rm e}$ = 100 cm$^{-3} $ to derive the theoretical H$\alpha$/H$\beta$, H$\gamma$/H$\beta$, and H$\delta$/H$\beta$ ratios (a typical approach for star-forming regions). We find that the difference between this approach and our iterative procedure are up to $\sim$0.04 dex, in particular, when $T_{\rm e}$ > 10000 K.
The uncertainties of the $E(B-V)$ values were estimated using a Monte Carlo simulation, generating 500 random values and assuming a Gaussian distribution with a sigma equal to the uncertainty of the associated Balmer ratio. 
The final adopted value of $E(B-V)$ for each spectrum is the weighted mean after discarding any negative $E(B-V)$.

In table \ref{tab:Evb}, we list the galaxy ID, the observational data, the $E(B-V)$ values for each Balmer line ratio (Columns 3--5), and the adopted $E(B-V)$ value (Column 6).
In Section~\ref{sub_sec:discussion}, we discuss the differences in $E(B-V)$ implied by the different Balmer lines with respect to H$\beta$ and other physical properties of CLASSY galaxies.

Finally, emission lines relative to H$\beta$ ($I_{\lambda}/I_{H_\beta}$) were corrected for reddening using the final $E(B-V)$ values reported in Table~\ref{tab:Evb} and the reddening function $f({\lambda})$ normalized to H$\beta$ by \citet{cardelli89}.
Note that at  optical wavelengths, the variation of extinction with \W\, are small \citep{shivaei:20, reddy16}. Therefore, our results are not affected by our choice of the extinction law selected for this sample.
The final errors are the result of adding in quadrature the uncertainties in the measurement of the fluxes and the error associated with the fits. 
We report the resulting emission line intensities for optical spectra analyzed in this work in Table~\ref{tab:LBT_intensities}.
Note that the optical emission lines are reported for the entire CLASSY sample in \citet{mingozzi22}, but only for a single spectrum per galaxy.
Given the comparison of multiple spectra per galaxy in this work, we have remeasured the emission lines of spectra in \citet{mingozzi22} for consistency.

\begin{deluxetable*}{c c c c c c c}
\tablewidth{0pt}
\caption{ Deredened emission line intensities measured for the LBT/MODS spectra for five CLASSY galaxies.
\label{tab:LBT_intensities}}
\tablehead{
Wavelength & Ion & J0808+3948 & J0944-0038 & J1148+2546   & J1323-0132 & J1545+0858 \\
 (\AA)          &     &       & SB\,2 & SB\,182 &       &        
}
\startdata
3726.04             &    [\ion{O}{2}]           &   $ 42.6\pm 3.2$        &   $ 36.0 \pm 2.7 $  &       $46.6 \pm 3.2$     &  $ 9.4 \pm 1.0 $     & $ 26.2 \pm 1.6 $  \\
3728.80             &    [\ion{O}{2}]           &   $ 30.6 \pm 3.0 $      &   $ 39.2 \pm 2.3 $  &     $68.7 \pm 3.3$     &  $ 9.4 \pm 1.0 $     & $ 33.4 \pm 1.6 $  \\
4101.71             &    H$\delta$              &   $ 28.42 \pm 1.43 $    &   $ 25.22 \pm 0.41 $  &    \nodata            &  $ 25.17 \pm 0.32$  & $ 24.5 \pm 0.28 $   \\
4340.44             &    H$\gamma$              &   $ 45.69 \pm 1.06 $    &   $ 46.61 \pm 0.69 $  &    45.2 $\pm$ 0.5    &  $ 46.26 \pm 0.49 $  & $ 45.62 \pm 0.42 $  \\
4363.21             &    [\ion{O}{ 3}]          &   \nodata               &   $ 12.42 \pm 0.43 $  &     10.2 $\pm$ 0.3     &  $ 20.15 \pm 0.38 $  & $ 12.87 \pm 0.27 $  \\
4861.35             &    H$\beta$               &   $ 100.0 \pm 3.0 $     &   $ 100.0 \pm 2.0 $   &     100.0 $\pm$1.1    &  $ 100.0 \pm 1.0 $   & $ 100.0 \pm 1.0 $  \\
4958.61             &    [\ion{O}{3}]           &   $ 18.5 \pm 1.3 $      &   $ 198.0 \pm 5.1 $   &     208.7 $\pm$ 4.4    &  $ 245.4 \pm 4.1 $   & $ 193.1 \pm 3.3 $   \\
5006.84             &    [\ion{O}{3}]           &   $ 54.5 \pm 1.7 $      &   $ 589.2 \pm 7.8 $   &    624.8 $\pm$6.5    &  $ 728.6 \pm 6.2 $   & $ 571.4 \pm 4.8 $   \\
5754.64             &    [\ion{N}{2}]           &   $ 1.31 \pm 0.26 $     &   \nodata             &    \nodata            & \nodata              & \nodata      \\
6312.06             &    [\ion{O}{2}]           &   \nodata               &   $ 1.47 \pm 0.06 $   &     1.3 $\pm$0.1      &  $ 0.82 \pm 0.22 $   & $ 1.09 \pm 0.04 $   \\
6548.05             &    [\ion{N}{2}]           &   $ 65.52 \pm 2.43 $    &   $ 1.07 \pm 0.8 $    &    1.8$\pm$0.8      &  $ 0.42 \pm 0.6 $    & $ 0.97 \pm 0.51 $   \\
6562.79             &    H$\alpha$              &   $ 299.5 \pm 9.8 $     &   $ 269.0 \pm 4.7 $   &    271.7 $\pm$ 4.1    &  $ 266.9 \pm 3.2 $   & $ 260.9 \pm 2.9 $   \\
6583.45             &    [\ion{N}{2}]           &   $ 196.56 \pm 7.29 $   &   $ 3.22 \pm 2.41 $   &    5.5 $\pm$ 2.3       &  $ 1.25 \pm 1.81 $   & $ 2.91 \pm 1.52 $   \\
6716.44             &    [\ion{S}{2}]           &   $ 20.54 \pm 0.76 $    &   $ 6.84 \pm 0.16 $   &    11.3 $\pm$ 0.3     &  $ 1.75 \pm 0.33 $   & $ 5.65 \pm 0.26 $   \\
6730.82             &    [\ion{S}{2}]           &   $ 26.72 \pm 0.92 $    &   $ 5.37 \pm 0.14 $   &    8.7 $\pm$ 0.2     &  $ 1.55 \pm 0.33 $   & $ 4.41 \pm 0.26 $   \\
7319.92             &    [\ion{O}{2}]           &  \nodata                &   $ 1.39 \pm 0.05 $   &    \nodata            &  $ 0.44 \pm 0.04 $   & \nodata     \\
7339.79             &    [\ion{O}{2}]           &  \nodata                 &   $ 1.14 \pm 0.05 $   &    \nodata            &  $ 0.37 \pm 0.04 $   & \nodata     \\
9068.60             &    [\ion{S}{3}]           &   $ 29.65 \pm 1.55 $    &   $ 10.1 \pm 0.33 $   &    10.8 $\pm$ 0.5     &  $ 4.41 \pm 0.21 $   & $ 7.87 \pm 0.21 $   \\
9530.60             &    [\ion{S}{3}]           &   $ 73.23 \pm 3.84 $    &   $ 24.66 \pm 0.69 $  &     26.1 $\pm$ 0.9     &  $ 10.9 \pm 0.52 $   & $ 19.45 \pm 0.51 $  \\
\hline
$E(B-V)$            &   &    $0.202\pm0.020$    &  $0.170\pm0.009$       &   $0.093\pm0.010$     &   $0.146\pm0.008$      & $ 0.08\pm0.007 $   \\
$F_{H\beta}$        &   &    $38.2 \pm 0.8$     &   $190.1 \pm 1.7$    &   $70.1 \pm 0.60$    &   $109.3 \pm 0.7$   &  $349.1 \pm 2.3$ \\
\enddata
\tablecomments{Intensity ratios are reported with respect to $I$(H$\beta$) = 100, where the observed
H$\beta$ fluxes (in units of 10$^{-16}$ erg s$^{-1}$ cm$^{-2}$) are reported in the second to last row.
The color excess value, $E(B-V)$, used to reddening correct the line ratios
are listed in the last row.}
\end{deluxetable*}

\section{ Physical conditions and chemical abundances}\label{sec:abundances}

We use the PyNeb package (version 1.1.14) \citep{luridiana15} in Python  to calculate the physical conditions and chemical abundances. We use atomic data for a 5-level atom model \citep{derobertis87}. We use the transition probabilities of  \citet{fft04} for O$^{+}$, O$^{2+}$, N$^{+}$ and, S$^{+}$ and  of  \citet{Podobedova:2009} for S$^{2+}$. For the collision strengths we use \citet{Kisielius:2009} for O$^{+}$, \citet{aggarwal99} for O$^{2+}$,  \citet{Tayal:2011} for N$^{+}$, \citet{Tayal:2011} for S$^{+}$ and \citet{Hudson:2012} for S$^{2+}$. 

\subsection{Nebular Density and Temperature}
To compute the electron density ($n_{\rm{e}}$), we use [\ion{S}{2}]~\W6717/\W6731 for all the set of observations and [\ion{O}{2}]~\W3726/\W3727 for LBT spectra and for four SDSS spectra.
We calculate the electron temperature using the line intensity ratios: [\ion{O}{2}]~(\W\W3726,3729)/(\W\W7319,7320 + \W\W7330,7331)\footnote{
Due to low spectral resolution it is not possible to resolve these lines. 
Hereafter referred to as [\ion{O}{2}]~\W3727 and [\ion{O}{2}]~\W\W7320,7330.}, [\ion{N}{2}]~(\W\W6548,6584)/ \W5755, 
[\ion{S}{3}]~(\W\W9069,9532)/\W6312, and 
[\ion{O}{3}]~(\W\W4959,5007)/\W4363, when they are available. 
The [\ion{S}{3}]~\W\W9069,9532 and [\ion{O}{2}]~\W\W7320,7330 
ratios can be affected by telluric and absorption features \citep{stevenson+94}. 
We carefully check the different spectra for the presence of emission/absorption 
features that can affect those lines. 
For LBT spectra, we measured the flux of both [\ion{S}{3}]~\W9069 and [\ion{S}{3}]~\W9532 
and compared the observed ratio to the theoretical ratio of [\ion{S}{3}]~\W9532/\W9069=2.47 
from \texttt{PyNeb} test for contamination by absorption bands. 
If the observed ratio of [\ion{S}{3}]~\W9532/\W9069 is within its uncertainty of the 
theoretical ratio, we use both lines to compute $T_{\rm e}$([\ion{S}{3}]). 
For [\ion{S}{3}] ratios that are outside this range, we discarded [\ion{S}{3}]~\W9069 when 
the observed ratio is larger than the theoretical ratio, and discarded [\ion{S}{3}]~\W9532
when the theoretical one is lower following \citet{arellano-cordova2020b, Rogers21}. 
Since it was only possible to measure [\ion{S}{3}]~\W9069 for most of the other spectra, we used it with 
the [\ion{S}{3}] theoretical ratio to estimate [\ion{S}{3}]~\W9532.

We also inspected [\ion{O}{2}]~\W\W7320,7330 for possible contamination by telluric 
absorption bands \citep{stevenson+94}. 
For the observations considered here, absorption bands were not detected around the red [\ion{O}{2}] lines. 
Note that due to the redshift of these galaxies, some other lines can also be affected by absorption bands. 
The [\ion{S}{2}] \W6731 line of J1323-0132 of LBT is contaminated by a telluric absorption band, 
which might impact the results of $n_{\rm e}$ measurement (see Sec~\ref{sub_sec:discussion} for more discussion). 
Additionally, the strongest emission lines in the optical spectra are at risk of saturating. 
We, therefore, carried out a visual inspection of the [\ion{O}{3}] \W5007 line and also compared its flux to the 
theoretical ratio of [\ion{O}{3}] \W5007/\W4959 for our sample. 
If [\ion{O}{3}] \W5007 is saturated, we instead used [\ion{O}{3}] \W4959 to estimate [\ion{O}{3}] \W5007 
using the theoretical ratio of 2.89 from \texttt{PyNeb}. 

In Table~\ref{tab:neTe}, we present the results of $n_{\rm e}$ in columns 2-3 and $T_{\rm e}$ in columns 3-7 for the whole sample. The uncertainties were calculated using Monte Carlo simulations in the similar way as $E(B-V)$.

\begin{deluxetable*}{lCCCCCCCCCCCC} 
\setlength{\tabcolsep}{1pt}
\tablecaption{Physical conditions and the ionic and total oxygen abundances for SDSS, LBT, MUSE and ESI observations }
\label{tab:neTe} 
\tablewidth{0pt}
\tablehead{
Galaxy      & n$_e$([SII])  & n$_e$([OII])
            & T$_e$([OII]) & T$_e$([NII]) & T$_e$([SIII])  & T$_e$[OIII])
            & T$_e$(Low)          & T$_e$ (High)        
            & O$^{+}$/H$^{+}$     & O$^{2+}$/H$^{+}$ & 12+ \\
            & (cm$^{-3}$)   & (cm$^{-3}$)      
            & (K)                 & (K)                 & (K)                  & (K)                 
            & (K)                 & (K)                &        ($\times$ 10$^{5}$)             &  ($\times$ 10$^{5}$) & log(O/H)}
\startdata
\multicolumn{10}{c}{\bf APO/SDSS}\\ \hline
J0021+0052   & 90\pm40    &    300\pm80 & 11600\pm450   & 12500\pm2600 & \nodata       & 10800\pm500   & 11600\pm450   & 10800\pm500   & 3.7 \pm 0.7   &   12.9 \pm 2.4  &  8.22\pm0.06 \\
J0808+3948   & 1100\pm250 &    \nodata  & \nodata       & \nodata       & \nodata       & \nodata       & \nodata       & \nodata       & \nodata       & \nodata         &   \nodata    \\
J0942+3547   & 50\pm30    &    \nodata  & \nodata       & 17600\pm6200  & \nodata       & 12600\pm200   & 12000\pm200   & 12600\pm200   & 2.4 \pm 0.2   &   8.3 \pm 0.8   &  8.05\pm0.03 \\
J0944-0038   & 120\pm30   &    \nodata  & \nodata       & \nodata       & 14100\pm300   & 15700\pm200   & 14000\pm100  & 15700\pm200   & 1.00 \pm 0.01   &   5.4 \pm 0.2   &  7.81\pm0.02 \\
J1024+0524   & 80\pm30    &    180\pm50 & 13200\pm550   & \nodata       & \nodata       & 14300\pm200   & 13200\pm550   & 14300\pm200   & 1.5 \pm 0.1   &   6.5 \pm 0.4   &  7.90\pm0.02 \\
J1044+0353   & 240\pm100  &    \nodata  & \nodata       & \nodata       & 15400\pm1100  & 19200\pm300   & 16500\pm200   & 19200\pm300   & 0.20 \pm 0.01   &  2.6 \pm 0.1   &  7.44\pm0.02 \\
J1129+2034   & 90\pm30    &    \nodata  & \nodata       & \nodata       & 10400\pm200   & 10300\pm200   & 16100\pm200   & 10300\pm200   & 7.1 \pm 0.7   &   15.1 \pm 1.4  &  8.35\pm0.04 \\
J1132+1411   & 140\pm60   &    \nodata  & \nodata       & \nodata       & 15000\pm2000  & 17300\pm350   & 10200\pm100   & 17300\pm350   & 1.1 \pm 0.3   &   1.6 \pm 0.4   &  7.42\pm0.10 \\
J1148 +2546  & 100\pm30   &   140\pm100 & 12100\pm350   & \nodata       & \nodata       & 13900\pm100   & 12100\pm350   & 13900\pm100   & 1.8 \pm 0.1   &   8.1 \pm 0.6   &  8.00\pm0.02 \\
J1323-0132   & 690\pm380  &    \nodata  & \nodata       & \nodata       & \nodata       & 17000\pm250   & 14900\pm100   & 17000\pm250   & 0.20 \pm 0.01   &   6.0 \pm 1.1   &  7.80\pm0.17 \\
J1418+2102   & 60\pm40    &    \nodata  & \nodata       & \nodata       & 16200\pm400   & 17900\pm300   & 15600\pm200   & 17900\pm300   & 0.40 \pm 0.01   &   3.2 \pm 0.3   &  7.56\pm0.02 \\
J1545+0858   & 90\pm20    &    270\pm100  & 11800\pm500  & \nodata       & \nodata       & 16200\pm200   & 11800\pm500   & 16200\pm200   & 0.60 \pm 0.01   &   4.9 \pm 0.3   &  7.74\pm0.02 \\
\hline\\
\multicolumn{10}{c}{\bf LBT/MODS}\\ \hline
J0808+3948  & 1000\pm250 & 1470\pm600  & \nodata       & 7700\pm1150   & \nodata       & \nodata       & 7700\pm1150   & 6500\pm700    & 46.8 \pm 22.6 &   11.5 \pm 5.6  &  8.77\pm0.24  \\
J0944-0038  & 140\pm10   & 480\pm170  & 15100\pm900   & \nodata       & 14900\pm400   & 15500\pm300   & 15100\pm100  & 15500\pm300   & 0.62 \pm 0.08   &   6.0 \pm 0.3   &  7.82\pm0.03  \\
J1044+0353  & 200\pm40   & $<100$  & 18200\pm1400       & \nodata       & 17700\pm500   & 19200\pm200   & 18200\pm1400    & 19200\pm200   & 0.13 \pm 0.01   &   2.6 \pm 0.2   &  7.44\pm0.02  \\
J1148+2546  & 110\pm50   & 100\pm10  & \nodata       & \nodata       & 13700\pm700   & 13800\pm200   & 12700\pm100   & 13800\pm200   & 1.00 \pm 0.10  &  8.3 \pm 0.6  &  7.97\pm0.02  \\
J1323-0132  & 640\pm650  & 680\pm390  & 15200\pm2600  & \nodata       & 17600\pm3800  & 17700\pm200   & 15200\pm2600  & 17700\pm200   & 0.10 \pm 0.01   &   5.2 \pm 1.2   &  7.74\pm0.02  \\
J1418+2102  & 80\pm40    &130\pm30  & 15200\pm850   & \nodata       & 14900\pm400   & 17800\pm200   & 15200\pm850   & 17800\pm200   & 0.30 \pm 0.01   &   3.6 \pm 0.3   &  7.60\pm0.02  \\
J1545+0858  & 130\pm30   &  190\pm100  & \nodata       & \nodata       & \nodata       & 16000\pm200   & 14200\pm100   & 16000\pm200   & 0.60 \pm 0.01   &   5.4 \pm 0.2   &  7.78\pm0.02  \\
\hline\\
\multicolumn{10}{c}{\bf VLT/MUSE}\\ \hline
J0021+0052  & 120\pm40 & \nodata    & \nodata       & 14100\pm1600  & \nodata       & 10600\pm400   & 14100\pm1600  & 10600\pm400   & 1.2 \pm 0.6   &   13.2 \pm 6.4 & 8.16\pm0.09  \\
J1044+0353  & 170\pm40 & \nodata    & \nodata       & \nodata       & 21500\pm900   & \nodata       & 17400\pm600   & 23800\pm1000  & 0.10 \pm 0.01   &   1.7 \pm 0.2  & 7.28\pm0.04  \\
J1418+2102  & 50\pm30  & \nodata    & \nodata       & \nodata       & 16700\pm400   & \nodata       & 14200\pm300   & 18100\pm500   & 0.60 \pm 0.10   &   3.3 \pm 0.3  & 7.59\pm0.03   \\  
\hline\\
\multicolumn{10}{c}{\bf Keck/ESI}\\ \hline
J0942+3547  & <100 & \nodata          & \nodata       & 10200\pm1300  & 11880\pm200   & 12400\pm200   & 11700\pm100   & 12400\pm200   &  3.5 \pm 0.3   &   9.1 \pm 0.8   &  8.10\pm0.30 \\
J0944-0038  & 100\pm10  & \nodata     & \nodata       & 11000\pm1900  & 15000\pm200   & 16700\pm200   & 14700\pm100   & 16700\pm200   &  0.80 \pm 0.01   &   4.5 \pm 0.1   &  7.73\pm0.01 \\ 
J1024+0524  & 100   & \nodata         & \nodata       & 12500\pm2600  & 18600\pm1500  & \nodata       & 15500\pm400   & 20400\pm700   &  0.5 \pm 0.1   &   3.1 \pm 0.1   &  7.55\pm0.04 \\
J1129+2034  & 90\pm20 & \nodata       & \nodata       & 10200\pm500   & 10300\pm100   & 9700\pm100    & 10200\pm500   & 10300\pm200   &  8.9 \pm 1.0   &   15.5 \pm 1.4  &  8.39\pm0.05 \\ 
J1132+5722  & 60\pm40 & \nodata       & \nodata       & \nodata       & 15800\pm1800  & 15800\pm1500  & 14200\pm1200  & 15800\pm1500  &  1.1 \pm 0.4   &   2.0 \pm 0.7   &  7.50\pm0.13 \\ 
J1148+2546  & 130\pm20  & \nodata     & \nodata       & 10800\pm1400  & 14100\pm200   & 14200\pm200   & 12900\pm100   & 14200\pm200   &  1.5 \pm 0.1   &   7.9 \pm 0.2   &  7.97\pm0.01  \\
\enddata
\tablecomments{For the ionic and total abundances we have used $T_{\rm e}$(Low) and $T_{\rm e}$(high), see text in Sec.~\ref{sec:abundances}. The ionic abundances, X$^{i+}$/H$^{+}$, and the total oxygen abundance in units of 12+log(O/H).}
\end{deluxetable*}
In \ion{H}{2} regions, a temperature gradient is expected 
in the interior of the nebula, associated with its different 
ionization zones (low-, intermediate-, and high-ionization) 
based on the ionization potential energy (eV) of the ions 
present in the gas \citep{osterbrock89, osterbrock06}.
To characterize the temperature structure of the nebular gas, 
we use $T_{\rm e}$([\ion{O}{3}]) as representative of the high-ionization 
region and $T_{\rm e}$([\ion{O}{2}]) or $T_{\rm e}$([\ion{N}{2}]) as 
representative of the low-ionization region. 
If both $T_{\rm e}$([\ion{O}{2}]) and $T_{\rm e}$([\ion{N}{2}]) 
are measured, we prioritize $T_{\rm e}$([\ion{O}{2}]) to characterize 
the low ionization region. 
It is because at low metallicity the [\ion{N}{2}] \W5755 line is too faint and the 
uncertainties associated to $T_{\rm e}$ are large in comparison to $T_{\rm e}$([\ion{O}{2}]) 
(see Table~\ref{tab:neTe}). When $T_{\rm e}$([\ion{O}{2}]), $T_{\rm e}$([\ion{N}{2}]) 
or $T_{\rm e}$([\ion{O}{3}]) is not detected, we 
use the temperature relation of \citet{garnett92} to estimate those temperatures, 
which are based on photoionization models: 

\begin{equation}
\label{eq:t2-t2}
\rm
T_{\rm e}[O\,II] \approx T_{\rm e}[N\,II] = 0.7 T_{\rm e}[O\,III] + 3000\,{\rm K},
\end{equation}

and, 

\begin{equation}
\label{eq:to3-ts3}
\rm
T_{\rm e}[\rm{S\,III}] = 0.83 T_{\rm e}[O\,III] + 1700\,{\rm K}.
\end{equation}

\begin{figure*}
\begin{center}
    \includegraphics[width=0.98\textwidth, trim=30 0 30 0,  clip=yes]{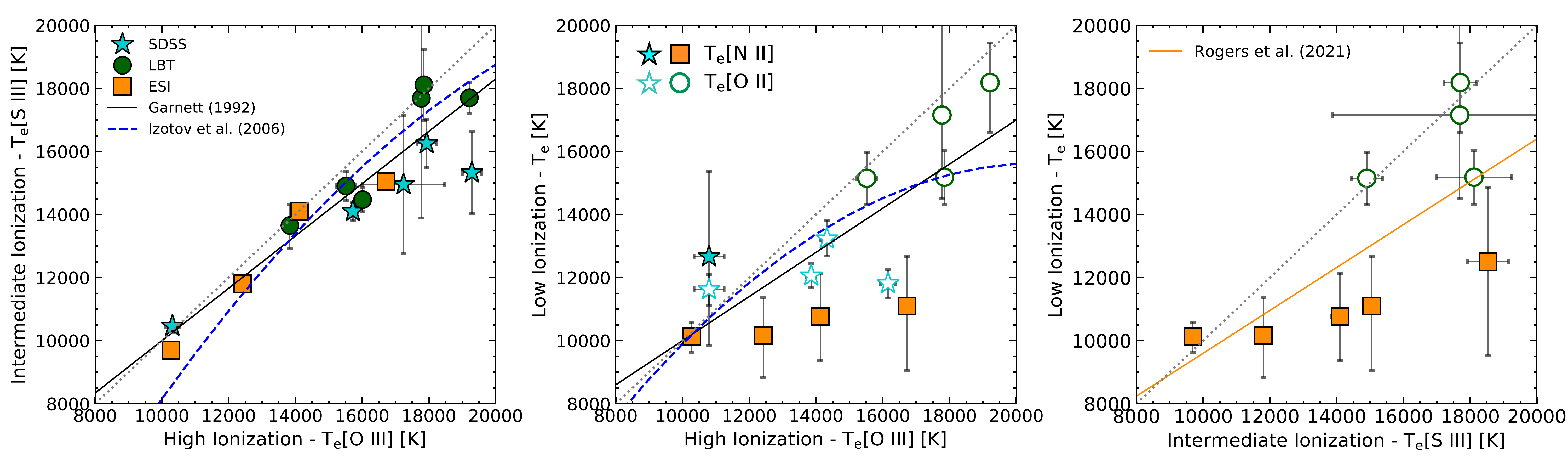}
    
    \caption{\textit{Left}: The $T_{\rm e}$([\ion{O}{3}]) - $T_{\rm e}$([\ion{S}{3}]) relation for some galaxies with SDSS, LBT and ESI spectra. The multiple apertures are labeled at the top of the left panel. The solid and dashed lines represent the temperature relations from \citet{garnett92} and \citet{izotov06} (low metallicity regime). \textit{Middle:} The $T_{\rm e}$([\ion{O}{3}]) - $T_{\rm e}$([\ion{N}{2}]) relation for ESI and SDSS sample indicated with solid symbols and the $T_{\rm e}$([\ion{O}{3}]) - $T_{\rm e}$([\ion{O}{2}]) relation for LBT and SDSS sample represented with empty symbols. \textit{Right:} The $T_{\rm e}$([\ion{S}{3}]) - $T_{\rm e}$([\ion{N}{2}]) relatios for ESI  results represented with solid symbols and the $T_{\rm e}$([\ion{S}{3}]) - $T_{\rm e}$([\ion{O}{2}]) relations for LBT results showed with empty symbols. The solid orange line represents the temperature relation derived by \citet{Rogers21}. The dotted lines in each panel represent the 1:1 relation.}
\label{fig:t-t_relation}
\end{center}
\end{figure*}

For those cases where the only available temperature is $T_{\rm e}$([\ion{S}{3}]), J1014 and J1418 of MUSE, we use the empirical relation from \citet{Rogers21} to estimate $T_{\rm e}$([\ion{O}{2}]): 

\begin{equation}
\label{eq:t2-ts}
T_{\rm e}[\rm{O\,II}] \approx T_{\rm e}[N\,II] = 0.68 T_{\rm e}[\rm{S\,III}] + 2800\,{\rm K}.
\end{equation}
Fig.~\ref{fig:t-t_relation} shows different temperature relation implied by the multiple apertures of CLASSY galaxies. In the left panel of this figure, we show the $T_{\rm e}$([\ion{O}{3}])-$T_{\rm e}$([\ion{S}{3}]) relation. The temperature relationships based on photoionization models of \citet{garnett92} and \citet{izotov06} (for low metallicity regime) are indicated by solid and dashed lines, respectively. In principle, the sample of galaxies follows both temperature relations, in particular, the LBT and ESI sample. However, at low temperatures ($T_{\rm e} < 14000$ K), the relationship of \citet{garnett92} is consistent with our results for ESI and SDSS observations (although few galaxies are present in these low temperature regime). As such, we feel confident in using the temperature relation of \citet{garnett92} to estimate  $T_{\rm e}$([\ion{O}{3}]) from $T_{\rm e}$([\ion{S}{3}]) when a unique $T_{\rm e}$ is available.

Low ionization temperatures, $T_{\rm e}$([\ion{N}{2}]) and $T_{\rm e}$([\ion{O}{2}]) are available for few CLASSY galaxies in this sample. However, in the middle and right panels of Fig.~\ref{fig:t-t_relation}, we present the low ionization temperatures versus $T_{\rm e}$([\ion{O}{3}]) and $T_{\rm e}$([\ion{S}{3}]), respectively. The solid and empty symbols represent the results for $T_{\rm e}$([\ion{N}{2}]) and $T_{\rm e}$([\ion{O}{2}]). For J0021+0051, it was possible to calculate both $T_{\rm e}$([\ion{N}{2}]) and $T_{\rm e}$([\ion{O}{2}]) using the SDSS spectrum (see also Table~\ref{tab:neTe}). We find that $T_{\rm e}$([\ion{N}{2}]) and $T_{\rm e}$([\ion{O}{2}]) show a difference of 900 K. However, the uncertainty derived to $T_{\rm e}$([\ion{N}{2}]) is large with respect to $T_{\rm e}$([\ion{O}{2}]). Such an uncertainty might be associated mainly to the measurement of the faint [\ion{N}{2}] \W5755 auroral in the SDSS spectrum.
On the other hand, $T_{\rm e}$([\ion{N}{2}]) and $T_{\rm e}$([\ion{O}{2}]) are expected to be similar due to that those temperatures are representative of the low ionization zone of the nebula (see Eqs~\ref{eq:t2-t2} and~\ref{eq:t2-ts}).
Previous works have studied the behaviour of the $T_{\rm e}$([\ion{N}{2}]) and $T_{\rm e}$([\ion{O}{2}]) relation using observations of \ion{H}{2} regions and star-forming galaxies \citep[ see e.g.,][]{izotov06, perezmontero07, croxall16, Rogers21, zurita21}.
Such results show a discrepancy and large dispersion between $T_{\rm e}$([\ion{N}{2}]) and $T_{\rm e}$([\ion{O}{2}]), whose origin might be associated to the measurements of the [\ion{N}{2}] \W5755 auroral in the case of $T_{\rm e}$([\ion{N}{2}]). For $T_{\rm e}$([\ion{O}{2}]), the [\ion{O}{2}]~\W\W7320, 7330 auroral lines  have a small contribution of recombination, depend on the electron density and reddening, and those lines can be contaminated by telluric lines ~\citep[e.g.,][]{stasinska05, perezmontero07, Rogers21,  arellano-cordova2020b}. Therefore, the impact of using $T_{\rm e}$([\ion{O}{2}]) in the determination of $O^{+}$ and the total oxygen abundance might be more uncertain in low ionization objects, i.e., high metallicity environments (12+log(O/H) $\gtrsim 8.2$.)
In Fig.~\ref{fig:t-t_relation}, we find that those results for $T_{\rm e}$([\ion{O}{2}]) (empty symbols) follow the relationships of \citet{garnett92} and \citet{izotov06}. As such, we also use \citet{garnett92} to estimate the low ionization temperature, $T_{\rm e}$([\ion{O}{2}]), when this temperature is not available.

On the other hand, the right panel of Fig.~\ref{fig:t-t_relation} also shows the $T_{\rm e}$([\ion{N}{2}]) - $T_{\rm e}$([\ion{S}{3}]) relation for ESI and LBT ($T_{\rm e}$([\ion{O}{2}])) observations. Such a relationship was previously presented for local star-forming regions by \citet{croxall16, berg20, Rogers21} showing a tight relation between those temperatures. The LBT and ESI observations follow the relationship derived by \citet{Rogers21} with some dispersion resulting mainly by the use of $T_e$([\ion{N}{2}]) results. We stress again that $T_ e$([\ion{N}{2}]) are uncertain in the high-$T_{\rm e}$/low metallicity limits because [\ion{N}{2}] \W5755 is weaker at such metallicities, for that reason we are more confident $T_{\rm e}$([\ion{O}{2}]) given the general agreement with the temperature relation presented in Fig.~\ref{fig:t-t_relation}. 
A future analysis of such relationships between $T_{\rm e}$([\ion{O}{2}]) and  $T_{\rm e}$([\ion{O}{3}]) and $T_{\rm e}$([\ion{S}{3}]) will be presented in Arellano-Cordova et al. (in prep.) using CLASSY galaxies, which cover a wider range of physical properties than local \ion{H}{2} regions \citep{croxall16, arellano-cordova20, berg20, Rogers21}. In addition, we will use photoionization models to constrain the temperature relations. 

\subsection{O/H Abundance}
The ionic abundances of O$^{+}$ were calculated using the adopted $T_{\rm e}$([\ion{O}{2}]) or $T_{\rm e}$([\ion{N}{2}]) representative for low ionization zone. To derive O$^{+}$, we use [\ion{O}{2}] \W3727 when this line was available and the [\ion{O}{2}] \W\W7320,7330 lines for the rest of the observations. For those spectra with both  [\ion{O}{2}] \W3727 and \W\W7320,7330 lines, we have compared the results obtained for O$^{+}$. We find an excellent agreements with small differences of 0.02 dex.   

For high ionization ions, O$^{2+}$, we use $T_{\rm e}$([\ion{O}{3}]) as representative of such an emitting zone. 
In Table~\ref{tab:neTe}, we list the adopted $T_{\rm e}$ associated to the low and high ionization zones in columns 8 and 9 ($T_{\rm e}$([\ion{O}{2}]) (low) and $T_{\rm e}$([\ion{O}{3}]) (high). In the same  Table~\ref{tab:neTe}, we also list the ionic abundances of O in columns 10 and 11.

The total oxygen abundance was calculated by summing the contribution of O$^{+}$/H$^{+}$ + O$^{2+}$/H$^{+}$. We neglected any contribution of O$^{3+}$/H$^{+}$ to O/H at it is negligible \citep{berg21a}. %
The uncertainties were calculated using Monte Carlo simulations by generating 500 random values assuming a Gaussian distribution with sigma associated with the uncertainty. In Table~\ref{tab:neTe}, we list the results of 12+log(O/H) in column 12 for the whole sample. In addition, in Table~\ref{tab:metal}, we present the results of metallicity for the multiple apertures in columns 2-5, and the differences in metallicity, $\Delta {\rm O/H} = {\rm O/H}_{\rm IFU,\,long-slit} - {\rm O/H}_{\rm SDSS}$ in columns 6-8.
%

\begin{deluxetable*}{CCCCCCCC }
\tablecaption{Metallicity Differences of CLASSY Spectra from Different Apertures \label{tab:metal}  }
\tablehead{
          & $\rm{SDSS}$        & $\rm{MUSE}$   & $\rm{LBT }$     & $\rm{ESI }$     & \multicolumn{3}{c}{} \\
Galaxy     &$\rm{(3'' cir.)}$ & $\rm{(2.5'' cir.)}$ & $\rm{(1'' slit)}$ & $\rm{(1'' slit)}$ & \multicolumn{3}{c}{$\Delta$$\rm{log O/H}$} } 
\startdata
J0021+0052   & 8.22$\pm$0.06  & 8.16$\pm$0.09      & \nodata             & \nodata             & -0.06 &       &       \\
J0942+3547   & 8.05$\pm$0.03  & \nodata            & \nodata             & 8.10$\pm$0.30   &       &       & +0.05 \\
J0944-0038   & 7.81$\pm$0.02  & \nodata            & 7.82$\pm$0.03       & 7.73$\pm$0.01   &       & +0.01 & -0.08 \\
J1024+0524   & 7.90$\pm$0.02  & \nodata            & \nodata             & 7.55$\pm$0.04   &       &       & -0.35 \\
J1044+0524   & 7.44$\pm$0.02   & 7.28$\pm$0.04      & 7.44$\pm$0.02       & \nodata             & -0.16 &  0.00 &       \\  
J1129+2034   & 8.35$\pm$0.04  & \nodata            & \nodata             & 8.39$\pm$0.05   &       &       & -0.04 \\
J1132+5722   & 7.42$\pm$0.10  & \nodata            & \nodata             & 7.50$\pm$0.13   &       &       & +0.08 \\
J1148+2546   & 8.00$\pm$0.02  & \nodata            & 7.97$\pm$0.02       & 7.97$\pm$0.01   &       & -0.03 & -0.03 \\   
J1323-0132   & 7.80$\pm$0.17  & \nodata            & 7.74$\pm$0.02       & \nodata            &       & -0.06 &       \\
J1418+2102   & 7.56$\pm$0.02  & 7.59$\pm$0.03       & 7.60$\pm$0.02      & \nodata            & +0.03 & +0.04 &       \\
J1545+0858   & 7.74$\pm$0.02  & \nodata            & 7.78$\pm$0.02       & \nodata            &       & -0.03 &       \\
\enddata
\tablecomments{
Metallicity values of 12+log(O/H) measured from the optical spectra of CLASSY galaxies.
Columns 2--3 list the abundances derived from the 3\as\ and 2.5\as\ circular apertures of SDSS and MUSE spectra, respectively, and columns 4--5 list the abundances derived from the 1\as\ longslit apertures of LBT and ESI spectra, respectively.
The differences in O/H between the IFU (MUSE) and long-slit (LBT and ESI) apertures versus the
SDSS apertures are represented as 
$\Delta$O/H =  O/H$_{ IFU,\,long-slit}$ - O/H$_{SDSS}$.}
\end{deluxetable*}

\subsection{Ionization Parameter}
\label{sec:ion_par}

We define O$_{32}$ = [\ion{O}{3}] \W5007/[\ion{O}{2}] \W3727 as proxy of the ionization parameter. We calculate O$_{32}$ for five galaxies of SDSS spectra (J0021+0052, J0808+3948, J1024+0524, J1148+2546 and J1545+0858) and for the whole LBT sample, whose [\ion{O}{2}] \W3727 line was available.

For the rest of the sample (ESI and MUSE spectra and some SDSS spectra), we obtained O$_{32}$ using the emissivities of [\ion{O}{2}] \W3727 and [\ion{O}{2}] \W\W7320,7330 using PyNeb. Such emissivities were calculated using  $T_{\rm e}$ (low) associated to the low ionization emitting zone (see Table~\ref{tab:neTe_differences}). We derived a representative factor between both [\ion{O}{2}] lines in the red and blue ranges for each galaxy. To estimate the intensity of [\ion{O}{2}] \W3727$_{\rm estimated}$, we used the observed emission lines of  [\ion{O}{2}] \W\W7320,7330 multiplied for the factor calculated from the emissivity ratio of $j$[\ion{O}{2}] \W3727/$j$[\ion{O}{2}] \W\W7320,7330, for each galaxy.  
In Table~\ref{tab:o32_log}, we list the results of O$_{32}$ for our sample of galaxies. We have identified those galaxies of SDSS sample  with estimates of [\ion{O}{2}] \W3727 with an asterisk, `$*$'.

\begin{deluxetable*}{CCCCCCCC }
\tablecaption{The Ionization Parameter Differences of CLASSY Spectra from Different Apertures \label{tab:o32_log}  }
\tablewidth{0pt}
\tablehead{
           & \rm{SDSS}        & \rm{MUSE}   & \rm{LBT }     & \rm{ESI }     & \multicolumn{3}{c}{} \\
Galaxy     &\rm{(3'' cir.)} &\rm{(2.5'' cir.)}&\rm{(1'' slit)}&\rm{(1'' slit)}& \multicolumn{3}{c}{$\Delta$\rm{log O$_{32}$}} } 
\startdata
J0021+0052  & $ 0.38\pm0.01^{*}$  &  $0.62\pm0.01$^{*} &   \nodata            &    \nodata         &  $+$0.24    &           &         \\
J0808+3948  & $-0.27\pm0.04$      &  \nodata       &   $-0.15  \pm  0.06$ &    \nodata         &           &   $+$0.12   &         \\
J0942+3547  & $ 0.56\pm0.01^{*}$  &  \nodata       &   \nodata            &    $0.36\pm0.01$   &           &           &  $+$0.20  \\
J0944-0038  & $ 0.76\pm0.01^{*}$  &  \nodata       &   $0.89  \pm  0.02$  &    $0.57\pm0.01$   &           &   $+$0.13   &  $-$0.19  \\
J1024+0524  & $ 0.65\pm0.01$      &  \nodata       &   \nodata            &    $0.65\pm0.04$   &           &           &   0.0   \\
J1044+0353  & $ 1.23\pm0.01^{*}$  &  $1.28\pm0.01$^{*} &   $1.23  \pm  0.01$  &    \nodata         &  $+$ 0.05    &   0.00  &         \\
J1129+2034  & $ 0.34\pm0.01^{*}$  &  \nodata       &   \nodata            &    $0.26\pm0.02$   &           &           &  $-$0.08  \\
J1132+5722  & $ 0.18\pm0.01^{*}$  &  \nodata       &   \nodata            &    $0.54\pm0.01$   &           &           &  $+$0.36  \\
J1148+2546  & $ 0.64\pm0.01$      &  \nodata       &   $0.73 \pm  0.02$   &    $0.64\pm0.01$   &           &   $+$ 0.09   &   0.0   \\
J1323-0132  & $ 1.65\pm0.01^{*}$  &  \nodata       &   $1.58 \pm  0.03$   &    \nodata         &           &   $-$0.07  &         \\
J1418+2102  & $ 0.96\pm0.01^{*}$  &  $0.98\pm0.01$^{*} &   $1.03 \pm  0.01$   &    \nodata         &  $+$0.02    &   $+$0.07   &         \\
J1545+0858  & $ 0.84\pm0.01$      &  \nodata       &   $0.98  \pm  0.02$  &    \nodata         &           &   $+$0.14   &         \\
\enddata
\tablecomments{
The ionization parameter (O$_{32}$ = [\ion{O}{3}] \W5007]/[\ion{O}{2}] \W3727) measured from the optical spectra of CLASSY galaxies.  Columns 2--3 list the results of log(O$_{32}$) derived from the 3\as\ and 2.5\as\ apertures of SDSS and MUSE spectra, respectively, and Columns 4--5 list the results of log(O$_{32}$) for the 1\as\ apertures of LBT and ESI spectra, respectively. The differences in log(O$_{32}$) between the IFU (MUSE) and long-slit (LBT and ESI) apertures versus the
SDSS apertures are represented as $\Delta$log O$_{32}$ = O$_{32\,(IFU,\,slit)}$ - O$_{32\,(SDSS)}$. $^{*}$ The intensity of [\ion{O}{2}] \W3727 was estimated using the [\ion{O}{2}]~\W\W7320, 7330 fluxes (see also Sec.~\ref{sec:ion_par}).} 
\end{deluxetable*}


\section{Discussion}
\label{sub_sec:discussion}
 In this section, we discuss the results implied for the different apertures obtained for CLASSY galaxies in Secs.~\ref{sec:ebv} and ~\ref{sec:abundances}, related to EW, ionization parameter, extinction, SFR, and metallicity.

\subsection{Equivalent widths aperture comparison} \label{sec:physical_properties}

We have determined the EWs of H$\alpha$ and [\ion{O}{3}] \W5007 and the ionization parameter for the the sub-sample of 12 CLASSY galaxies. In Table~\ref{tab:Evb}, we list the results for the EWs in columns 7-8.


The EWs of H$\alpha$ is ranging from 68 \AA\, to 1620 \AA\ (1.8-3.2 in dex) for SDSS and LBT, and from 280 \AA\, to 1800 \AA\ (2.4-3.3 in dex), for ESI and MUSE (see Table~\ref{tab:Evb}). Such values of EWs are expected in local star-forming galaxies \citep[e.g.,][]{berg16,senchyna19}. For EW([\ion{O}{3}]), we report values ranging between $\sim$ 7 \AA\, and 2100 \AA\ (0.8-3.3 in dex) for SDSS and LBT, from $\sim$ 280 \AA\, to 1800\AA\ (2.4-3.3 in dex) for ESI, and  $\sim$430 \AA\, to 1500 \AA\ (2.6-3.2 in dex) for MUSE. 

\begin{figure} 
\begin{center}
    \includegraphics[width=0.35\textwidth, trim=30 0 30 0,  clip=yes]{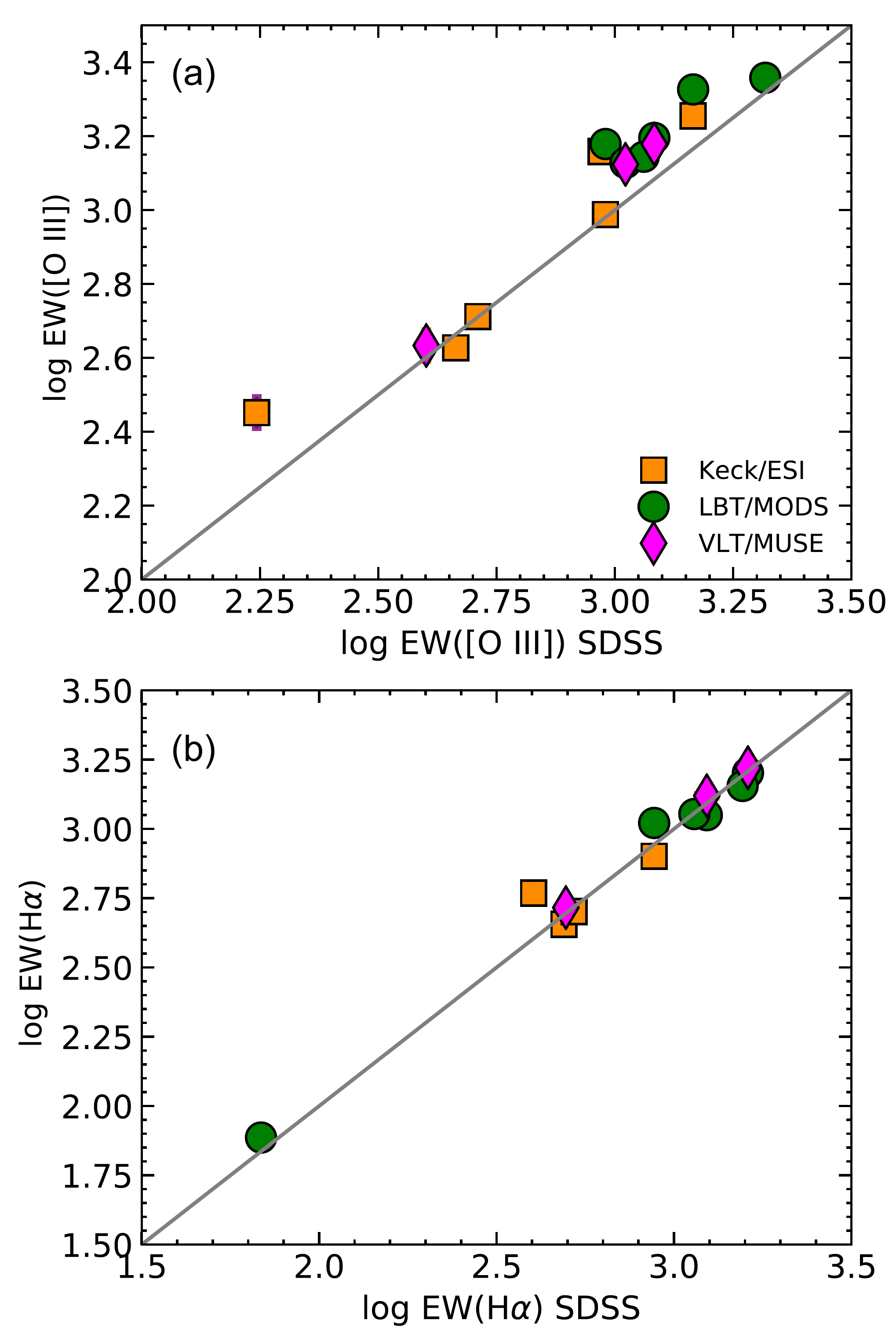}
    \caption{Comparison of log EW of [\ion{O}{3}]\ (a) and H$\alpha$ (b) of MUSE, LBT and ESI apertures with respect to SDSS. The solid line represents the 1:1 relationship. The results of the EWs are reported in Table~\ref{tab:Evb}.}
    \label{fig:ews_comparison}
  \end{center}
    \end{figure}

In panels (a) and (b) of Fig.~\ref{fig:ews_comparison}, we show the comparison between EWs of [\ion{O}{3}]\ and H$\alpha$ derived using SDSS spectra and those results using long-slit and IFU spectra of LBT, ESI and MUSE.
Overall, we find consistent results among the different setups with differences lower than $~$0.07 dex for EW(H$\alpha$) with respect to SDSS EW(H$\alpha$), except for J1132+5722 that shows a difference of 0.2 dex. However, for EW([\ion{O}{3}]), the differences can reach up to 0.2 dex for ESI and LBT spectra, and for MUSE the differences can reach up to 0.1 dex with respect to SDSS EW([\ion{O}{3}]).
These differences can be due to the different apertures collecting different relative amounts of nebular emission and stellar continuum.

We have also compared the results of EWs as a function of the fraction of total optical light captured by different aperture sizes used in these observations. We estimated the total galaxy radius using $r$-band images from Pan-STARRS following the same procedure as \citet{berg22} in CLASSY Paper I. 
We have approximated the total extent of the $r$-band light, $R_{100}$, as the radius containing nearly all of the integrated light (although 92.5\% was used to avoid unphysical sizes).
Because our targets have the majority of their light contained within a compact center, we define their galaxy sizes as $R_{100}/2$ (see also Fig.~\ref{fig:apertures}).

Using this value, we estimated the percentage of the light enclosed within the long-slit and IFU apertures as the fraction of the area covered by the respective aperture sizes relative to the adopted optical area of the galaxies: $A_{\rm aper.}$/$A_{R_{100/2}}$. 

\begin{figure*} 
\begin{center}
    \includegraphics[width=0.9\textwidth, trim=30 0 30 0,  clip=yes]{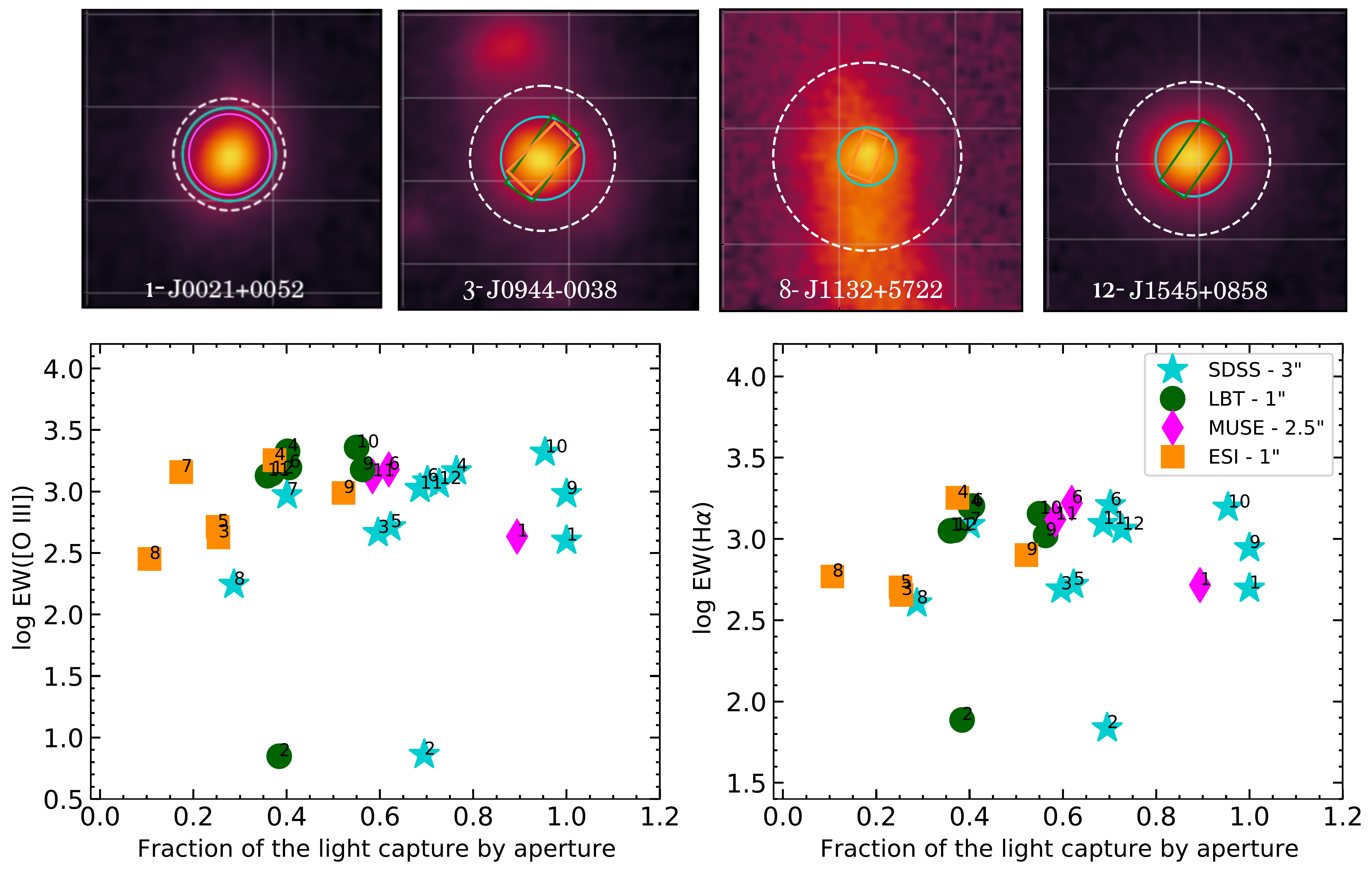}
    \caption{\textit{Top:} The SDSS $r$-band images of four galaxies of this sample (see also Fig.~\ref{fig:apertures}). The 2.5\arcsec\, MUSE aperture and the 1\arcsec\, LBT and ESI apertures are shown as a magenta circle and  green and orange boxes, respectively. The 3\arcsec SDSS aperture is also shown as a cyan circle, and the dotted circle represents the total galaxy radius defined as $R_{100}/2$ (see Table~\ref{tab:journal} and Sec.~\ref{sec:physical_properties}).
    \textit{Bottom:} Comparison of log EW of [\ion{O}{3}] (a) and H$\alpha$ (b) for the SDSS, MUSE, LBT, and ESI apertures versus the fraction of total optical light enclosed by the respective spectral aperture, $A_{aper}$/$A_{R_{100}/2}$. The results of the EWs are reported in Table~\ref{tab:Evb}. The numbers identify the ID number of the galaxy listed in Table~\ref{tab:journal}.}
    \label{fig:ews_r50}
  \end{center}
    \end{figure*}

The top panel of Fig.~\ref{fig:ews_r50} shows the SDSS $r$-band images for a representative sample of our CLASSY galaxies, which show compact and elongated morphologies. The different apertures used in these observations are overplotted as in Fig~\ref{fig:apertures}. The dashed circle represents the total galaxy radius, $R_{100}/2$, showing that most of the flux is captured by the IFU and LS apertures. On the other hand, the bottom panel of Fig.~\ref{fig:ews_r50} shows the comparison of the fraction of light ($A_{\rm aper.}$/$ A_{R_{100/2}} $) covered by different apertures as a function of the EWs of [\ion{O}{3}] \W5007 (a) and H$\alpha$ (b). We have identified the galaxies according to their numbers in Table~\ref{tab:journal}. The same number corresponds to the same galaxy in the different observations. Fig.~\ref{fig:ews_r50} indicates that although the light fraction changes up to $\sim$20\% between observations of a given galaxy, there is little change in the EWs: 
the difference in EW is lower than $\sim$0.1 dex for most of the galaxies (see, also, Fig.~\ref{fig:ews_comparison} and Fig.~\ref{fig:Ebv_vs_EW}).
Interestingly, this suggests that the stellar and nebular emission have similar distributions (e.g, centrally-dominant distributions) within these compact galaxies such that aperture differences have little affect on EW measurements.

We also find that in most galaxies, the percentage of light covered is larger than 60\% for apertures of  3\arcsec\ and 2\farcs5. For long-slit LBT and ESI spectra, the light enclosed within such aperture ranges between 20\% and 60\%. We also note that for J1129+2034 and J1132+5722, their percentage is lower than 20\% for either IFU ($\sim$ 30\%) of long-slit apertures. In fact, those galaxies have an elongated form (see top panel of Fig.~\ref{fig:ews_r50}). However, the observations for those galaxies are where the peak of star formation is located. 
 
\subsection{Ionization parameter comparison}
We have analyzed the variations of the ionization parameter with aperture size.
Overall, we find that differences in log(O$_{32}$) can reach values up to $\sim$0.14 dex. There are some exceptions with differences larger than 0.2 dex. Note that part of these large differences might be due to the estimate of [\ion{O}{2}] \W3727 using the [\ion{O}{2}] \W\W7320,7330 lines. To check this, we compare the values of O$_{32}$ for the galaxies of the SDSS and LBT sample for which the [\ion{O}{2}] \W3727 emission lines are available.
Such galaxies are J0808+3948, J1148+2546, and J1545+0858. We find that the differences calculated are up to 0.14 dex, except for J1148+2546 that shows a differences of 0.08 dex.  In this case, the LBT spectrum provides higher values of O$_{32}$ than the SDSS spectrum. Part of these differences might be due to the sampled region within the galaxy. However, we find the same result for J1148+2546 using the ESI spectrum (see Table~\ref{tab:o32_log}).  

\subsection{$E(B-V)$ aperture comparison}

\begin{figure} 
\begin{center}
\includegraphics[width=0.35\textwidth, trim=30 0 30 0,  clip=yes]{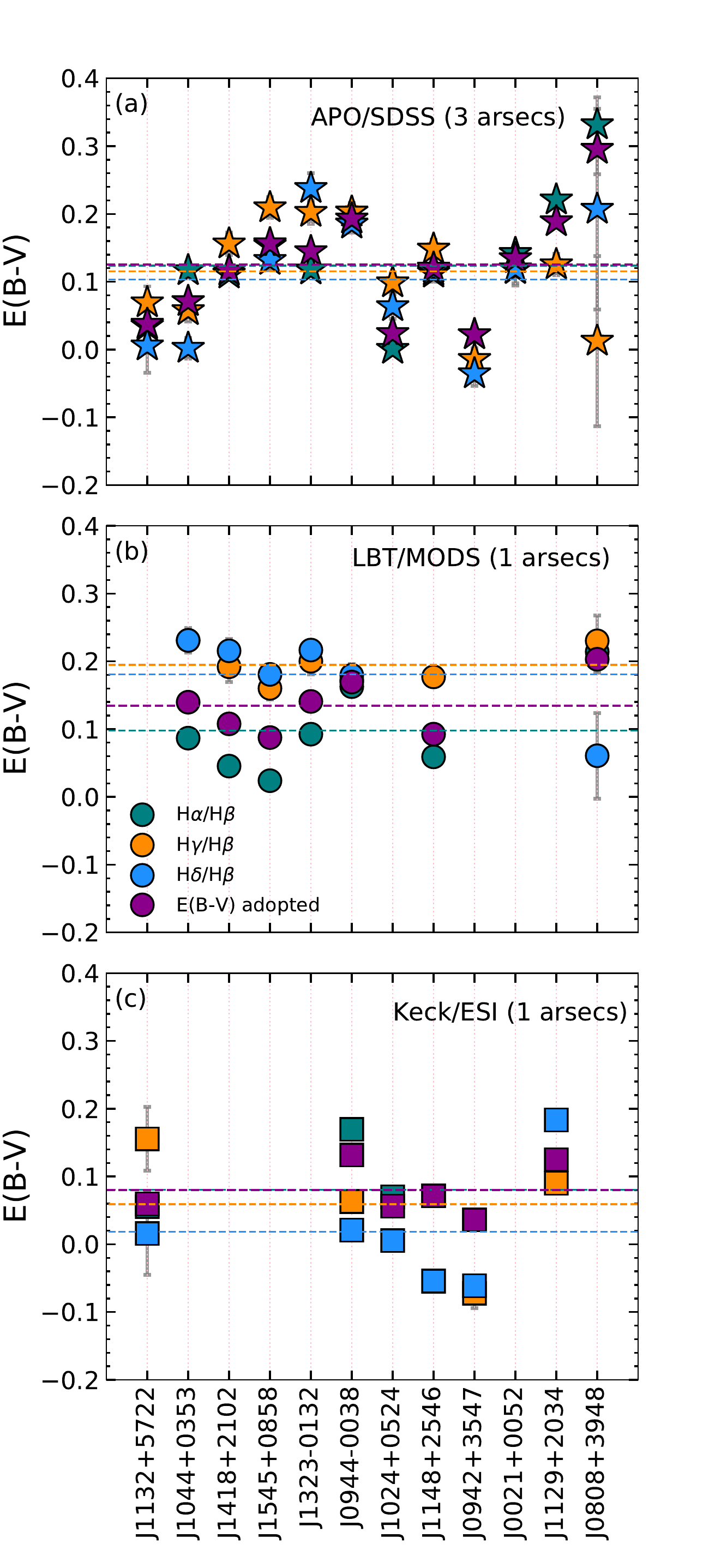}

    \caption{Comparison $E(B-V)$ derived using H$\alpha$/H$\beta$, H$\gamma$/H$\beta$ and H$\delta$/H$\beta$ for SDSS (a) LBT (b), and ESI (c) spectra, where the symbols represent each instrument, stars, circles, and squares, respectively. The galaxies are ordered at increasing metallicity, see Table~\ref{tab:metal}. The symbols indicate the $E(B-V)$ values derived using the Balmer decrement (teal, orange, and blue) and the final $E(B-V)$ values (purple, see Table~\ref{tab:Evb}). The dashed lines represent the mean values of $E(B-V)$ derived using the Balmer lines and the adopted $E(B-V)$ value for each observation (see Table~\ref{tab:dispersions}).}
    \label{fig:Ebv}
  \end{center}
    \end{figure}

The results presented in Table~\ref{tab:Evb} provide a broad picture of the variations derived using Balmer ratios to calculate the reddening of this sample of star-forming galaxies. In. Fig~\ref{fig:Ebv}, we show the behavior of H$\alpha$/H$\beta$, 
H$\gamma$/H$\beta$, and H$\delta$/H$\beta$ and the adopted value of $E(B-V)$ for SDSS, ESI, and LBT apertures. The $X$-axis shows the 12 galaxies studied here ordered by increasing metallicity. 
The different symbols indicate the Balmer ratio used to estimate $E(B-V)$ using Eq.~\ref{eq:ebv}. We discarded the results of MUSE because of the missing coverage of H$\gamma$/H$\beta$ and H$\delta$/H$\beta$ for J1044+0353 and J1418+2102. 
 
 In general, SDSS, ESI, and LBT observations show no variation distribution of $E(B-V)$ with respect to metallicity for the different estimates of $E(B-V)$ using Balmer lines. 
The dashed lines in the top panels of Fig.~\ref{fig:Ebv} indicate the mean values of $E(B-V)$ for each set of observations in color code with the symbols, including the adopted values of $E(B-V)$. Such values, mean, and dispersion for the results of $E(B-V)$ using  Balmer ratios are reported in Table~\ref{tab:dispersions}. Note those estimates of the dispersion of $E(B-V)$ in Table~\ref{tab:dispersions} depends on the number of galaxies with measurements of H$\alpha$/H$\beta$, H$\gamma$/H$\beta$, and H$\delta$/H$\alpha$~ available in the observed spectra. For the SDSS sample, we find similar values within the uncertainties implied for the different Balmer ratios used. 

In Fig.~\ref{fig:Ebv}, it is observed that the SDSS spectra with an aperture size of 3\arcsec show less scatter and lower uncertainties in the $E(B-V)$ implied by the different Balmer line ratios in comparison to those obtained using long-slit spectra. In addition, it seems LBT results show systematic effects for individual $E(B-V)$ results, but the adopted $E(B-V)$ is consistent with SDSS results.  

\begin{deluxetable}{l c c c} 
\tablecaption{Mean and dispersion values of E(B-V) using Balmer decrement for the different apertures.}
\label{tab:dispersions} 
\tablewidth{0pt}
\tablehead{
$E(B-V)$             &  SDSS               &  ESI                 & LBT }
\startdata
H$\alpha$/H$\beta$      &  $0.124\pm0.094$    & $0.080\pm0.052$     & $  0.098\pm 0.067$ \\
H$\gamma$/H$\beta$      &  $0.116\pm0.074$    & $0.059\pm0.095$     & $  0.195\pm 0.027$ \\
H$\delta$/H$\beta$      &  $0.103\pm0.093$    & $0.018\pm0.088$     & $  0.181\pm 0.062$ \\
\\
$E(B-V)$ & $0.125\pm0.080$ & $0.080\pm0.039$ & $ 0.134\pm 0.042$ \\
\enddata
\tablecomments{The mean and dispersion values to each Balmer ratio correspond to the total number of galaxies of each sample, and those are represented in Fig.~\ref{fig:Ebv}. The $E(B-V)$ shows the mean and dispersion value for the whole sample of observations in each instrument, SDSS, LBT, and ESI.}
\end{deluxetable}

For LBT spectra, $E(B-V)$ derived using H$\gamma$/H$\beta$ and H$\delta$/H$\beta$ show higher values with respect to H$\alpha$/H$\beta$ in comparison to SDSS and ESI. For example, such differences implied in the $E(B-V)$ estimate might be due to stellar continuum subtraction. In \ion{H}{2} regions of local disk galaxies, the discrepancy between H$\alpha$/H$\beta$ and  H$\gamma$/H$\beta$ and H$\delta$/H$\beta$ have been reported in different works, impliying that this issue is common in other environments \citep{croxall16, Rogers21}.  

On the other hand, some of the results of $E(B-V)$ provide negative reddening calculated for some spectra of SDSS and ESI (see Fig.~\ref{fig:Ebv} and Table~\ref{tab:Evb}), derived using H$\gamma$/H$\beta$ and H$\delta$/H$\beta$. Previously, \citet{kewley05}
found similar behavior for some galaxies of their sample. These authors pointed out as possible cause the errors in the stellar subtraction, flux calibrations and uncertain measurements due to noise. However, those values should agree within the errors. 
We have revised the emission lines fitting in those spectra with negative $E(B-V)$ around H$\gamma$ and H$\delta$. In general, the range in a wavelength where those lines are located shows no apparent issues in the flux measurements of those Balmer lines. However, we do not exclude that stellar continuum subtraction might affect the flux measured in those lines as was suggested by \citet{Groves12}. In an incoming paper, we will assess the impact of the reddening correction implied by Balmer and Paschen lines calculating $E(B-V)$ \citep{Rogers21, mendez-delgado21, aver21} using a different set of observations of CLASSY. In particular, the LBT observations of this paper (see Fig.~\ref{fig:spectralbt}), which cover a wide range in wavelength. It allows analyzing the in much more detail the reddening correction, dust geometry \citep{scarlata09b}, the ionization structure of the gas, the electron density and temperature using different diagnostics, chemical abundances of different ions and the very high ionization emission lines \citep{berg21a}.

\begin{figure} 
\begin{center}
    \includegraphics[width=0.35\textwidth, trim=30 0 30 0,  clip=yes]{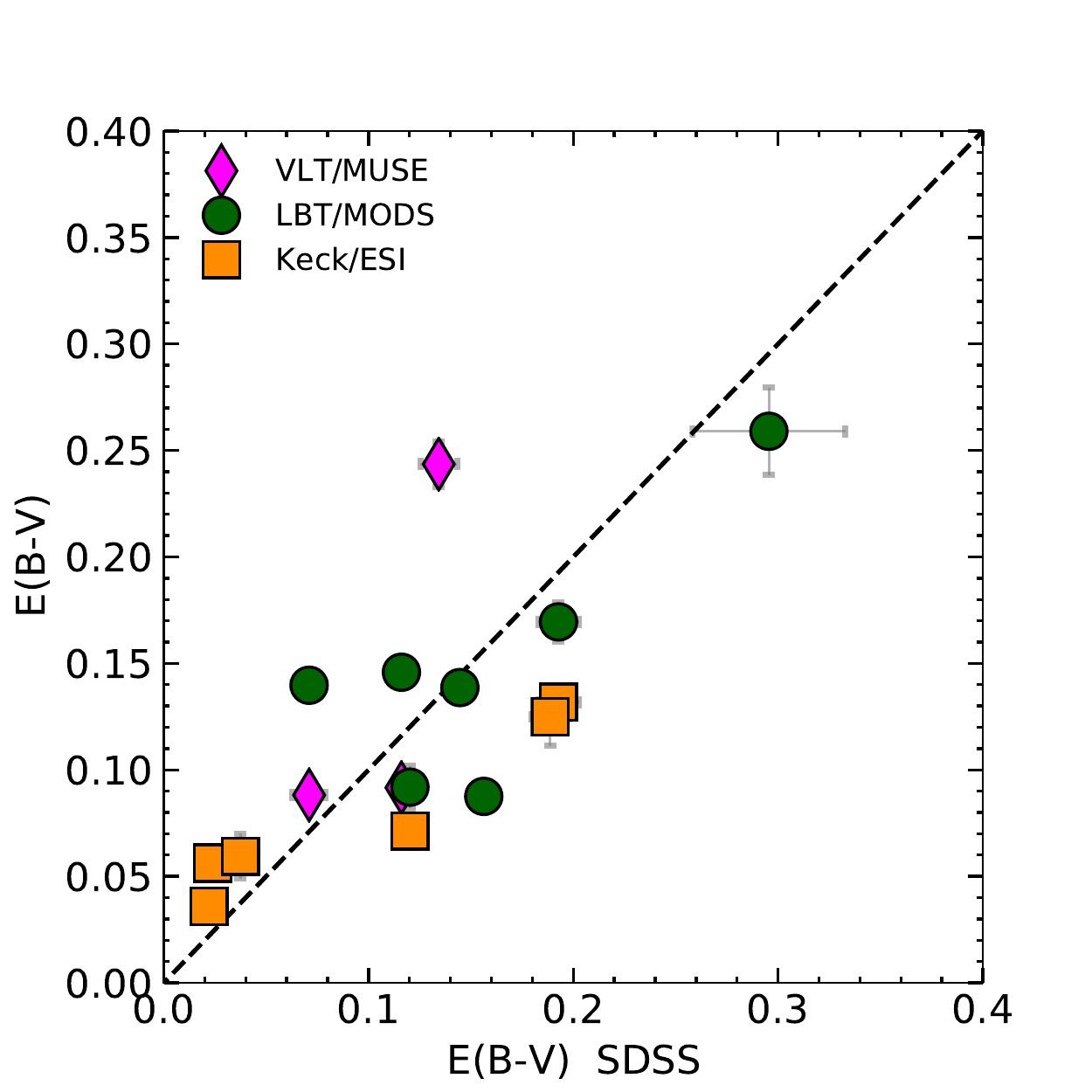}
    \caption{Comparison of the final $E(B-V)$ values derived from MUSE, ESI and LBT spectra with respect to the SDSS sample. The symbols indicate the different observations, IFU (VLT/MUSE) and long-slit (Keck/ESI and LBT/MODS) with respect to APO/SDSS sample. The dashed line represents the 1:1 relationship.}
    \label{fig:Ebv_final}
  \end{center}
    \end{figure}

 Now, we show the comparison of different apertures implied in the results of the adopted $E(B-V)$ with respect to the SDSS results, which it shown in Fig.~\ref{fig:Ebv_final}. The symbols represent the observations implied for the different apertures, and the dashed line indicates the 1:1 relationship. In general, the differences, $\Delta E(B-V) = E(B-V)_{\rm IFU,\, long-slit} -  E(B-V)_{\rm SDSS}$ are lower than 0.1 dex for most of the observations (see also Table~\ref{tab:Evb}).

\begin{figure*} 
\begin{center}
    \includegraphics[width=0.9\textwidth, trim=30 0 30 0,  clip=yes]{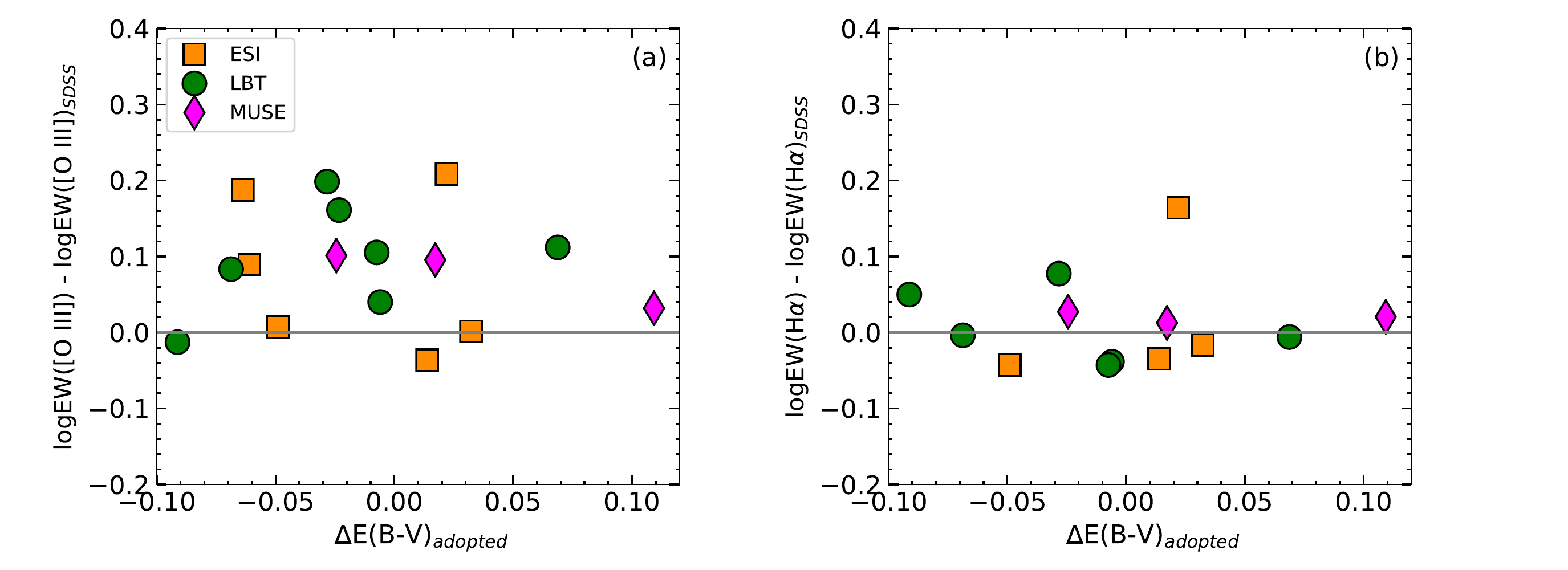}
    \caption{Color-excess differences, $
    \Delta E(B-V)_{\rm adopted} =  E(B-V)_{IFU, LS} - E(B-V)_{\rm SDSS}$ as a function of the adopted EW differences with respect to SDSS results for [\ion{O}{3}]~\W5007 (a) and H$\alpha$ (b). The symbols show the IFU (VLT/MUSE) and LS (LBT/MODS, Keck/ESI) observations, respectively. Note that the results for EWs(H$\alpha$) correspond to those spectra with measurements of H$\alpha$ (see Table~\ref{tab:Evb}).}
    \label{fig:Ebv_vs_EW}
  \end{center}
    \end{figure*}

Finally, we also compare the reddening derived by averaging the values from three different Balmer line ratios {E(B-V)$_{\rm adopted}$}) as a function [\ion{O}{3}]~\W5007 and H$\alpha$ EWs. In Fig.~\ref{fig:Ebv_vs_EW}, we show the comparison between the differences between the E(B-V) adopted values and the EWs of [\ion{O}{3}]~\W5007 (a) and H$\alpha$ (b) with respect to the SDSS results. We find that at least for the range implied for this sample, $E(B-V)$ versus EWs seems to be constant. Note that the $E(B-V)$ versus EW([\ion{O}{3}]) shows a larger dispersion than EW([H$\alpha$]), although for a different number of galaxies (see Table~\ref{tab:Evb}). However, the difference for LS and IFU sample are lower than 0.1 dex for most of the galaxies for either IFU or LS apertures.  

\subsection{Density, temperature, and metallicity comparison}
\label{sec:den_ten_metal}
Here, we analyze if the use of different apertures impacts the physical condition and chemical abundance determinations. In principle, both density and temperature diagnostics can be affected by observational problems such as the measurements of faint auroral lines ([\ion{O}{3}]~\W4363 or [\ion{N}{2}]~\W5755), uncertainties involved with extinction and flux calibration, telluric absorption among others \citep{arellano-cordova20}. We have carefully inspected each of the individual spectra implied for the different set of observations (see also Sec.~\ref{sec:abundances}). It is important to evaluate only the differences indicated for aperture effects.

\begin{figure*} 
\begin{center}
    \includegraphics[width=0.98\textwidth, trim=30 0 30 0,  clip=yes]{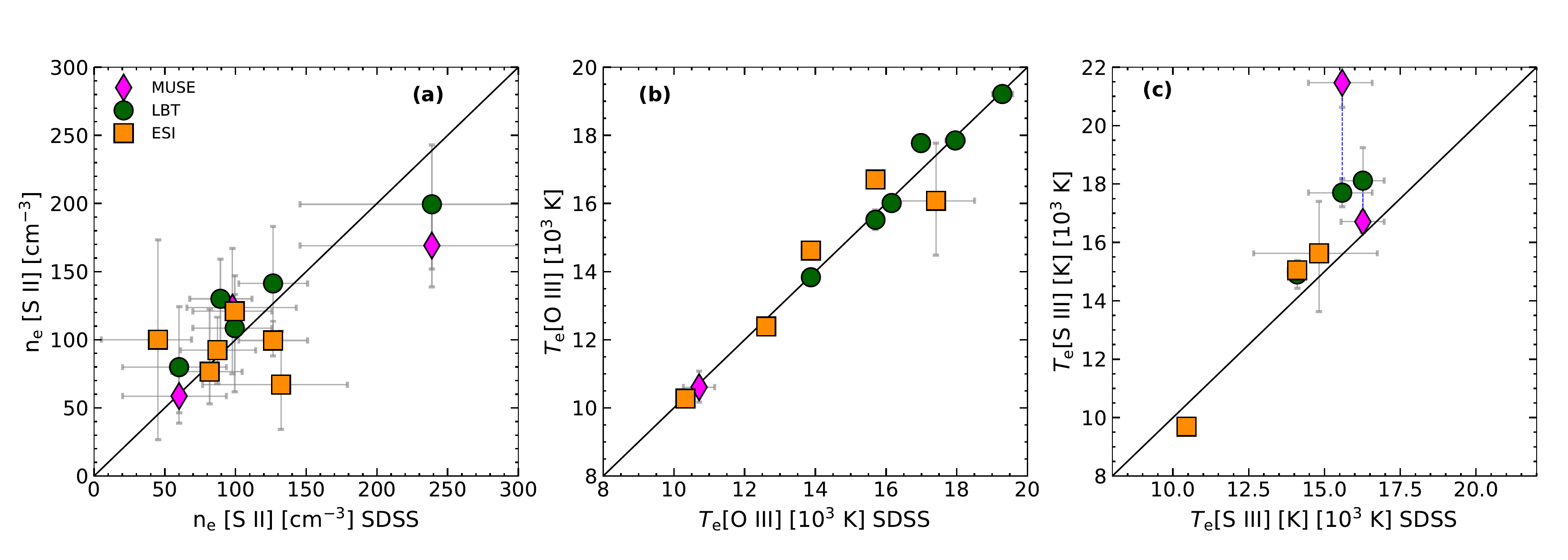}
    \caption{Electron density and temperature for the multiple apertures of CLASSY galaxies. (a) Electron density, $n_{\rm e}$[\ion{S}{2}]~comparison for the multiple observations. (b) Comparison of $T_{\rm e}$([\ion{O}{3}]) from SDSS spectra versus $T_{\rm e}$([\ion{O}{3}]) using multiple aperture spectra of MUSE, LBT and ESI. We discard those galaxies with high density, J0808, and J1323, to better compare those with low density. (c) $T_{\rm e}$([\ion{S}{3}]) calculated using SDSS spectra in comparison with $T_{\rm e}$([\ion{S}{3}]) calculated using multiple aperture spectra. The dotted line joins the same galaxy from MUSE and LBT data with a large difference in both SDSS and LBT data. The solid lines indicate the one-to-one relation. The different symbols correspond to the multiple apertures, which are labeled at the top of the left panel.}
    \label{fig:ne_te}
  \end{center}
    \end{figure*}

For density, it was possible to obtain measurements for all the multiple apertures using [\ion{S}{2}]~intensities (see Table~\ref{tab:neTe} and Sec.\ref{sec:abundances}). 
In principle, We use SDSS observations as a comparison sample for our results from the other instruments. The reason is that for SDSS, we can measure most of the physical properties of each galaxy, and because the aperture implied in SDSS (3\arcsec) sample is mapping a similar emission region to that of the HST/COS aperture (2.5\arcsec\,). 
For the SDSS sample, we obtain densities of 50 $< n_{\rm e}$ (cm$^{-3}) < 1100$. J0808+3948, J1044+0353, and J1323-0132 show high values of $n_{\rm e}$ > 300 cm$^{-3}$ in comparison with the rest of the sample. In panel (a) of Fig.~\ref{fig:ne_te}, we show the comparison of the results for $ n_{\rm e} < 300$ using SDSS spectra versus the densities calculated using the multiple spectra (we discard the high-density values for display purposes corresponding to J0808+3849 and J1323-0132). We find that at low density, $n_{\rm e}$ < 150 cm$^{-3}$, where more of the CLASSY galaxies are located, long-slit observations show consistent results within the uncertainties with the SDSS sample. For J0808+3948 and J1044+0353, with high-density values, LBT and SDSS observations show small differences of 40 and 100 cm$^{-3}$, respectively. 

In our first analysis of the electron density for J1323-0132 of LBT, we find a difference of $370$  cm$^{-3}$ in comparison to SDSS aperture. We note that the observed wavelengths of the [\ion{S}{2}] lines in J1323-0132 are affected by telluric absorption. We fitted those lines again, taking into account the absorption feature. For J1323-0132, we calculate a new value of $n_{\rm e}$ = 640$^{+800}_{-500}$ cm$^{-3}$, which implies a difference of $50$ cm$^{-3}$ with respect to the previous contaminated value. In Table~\ref{tab:neTe_differences}, we show the differences in density, temperature and metallicities of IFU and LS apertures concerning SDSS measurements. Those galaxies with multiple apertures using IFU and LS are also indicated in Table~\ref{tab:neTe_differences} for comparison (J1044+0353 and J1418+2102). In general, we find that $n_{\rm e}$[\ion{S}{2}] implied for the different apertures are in good agreement with those results of SDSS aperture with differences lower than 100 cm$^{-3}$.

On the other hand, we have calculated $n_{\rm e}$[ \ion{O}{2}] cm$^{-3}$ for some galaxies. It includes the whole sample of the LBT observations and four galaxies of SDSS (J0021+0052, J1024+0524, J1044+0353, and J1545+0858, see also Table~\ref{tab:neTe}). Firstly, we compare the results between the density diagnostics of $n_{\rm e}$[ \ion{O}{2}] and $n_{\rm e}$[ \ion{S}{2}] derived for the LBT spectra. Overall, we find differences lower than 100 cm$^{-3}$ for most of the LBT sample. The exception is J0021+0052 and J0808+3948 whose difference is up 470 cm$^{-3}$. In similar way, the four galaxies with $n_{\rm e}$[ \ion{O}{2}] show differences of up to $\sim$ 210 cm$^{-3}$. Note that the density diagnostic of $n_{\rm e}$[ \ion{O}{2}] provides values higher than the results implied by $n_{\rm e}$[ \ion{S}{2}] in both samples. Next, we compare the results implied by $n_{\rm e}$[ \ion{O}{2}] for two galaxies in common in the LBT and SDSS sample (J1148+2546 and J1548+0858). We find differences lower than 80 cm$^{-3}$.

Concerning the electron temperature, we have calculated $T_{\rm e}$([\ion{O}{3}]) and $T_{\rm e}$([\ion{S}{3}]) for most of the spectra of CLASSY galaxies (see Table~\ref{tab:neTe}). $T_{\rm e}$([\ion{O}{3}]) was measured in the whole SDSS sample (with the exception of J0808+3948) and in some galaxies for the rest of the observations with different apertures. In panel (b) of Fig.~\ref{fig:ne_te}, we show the comparison of $T_{\rm e}$([\ion{O}{3}]) results for the multiple apertures. In general, $T_{\rm e}$([\ion{O}{3}]) for IFU and long-slit apertures are consistent with SDSS results with differences lower than 700 K for most of the objects (see Table \ref{tab:neTe_differences}). J1132+5722 of ESI sample show a difference of  $-1500$~K in $T_{\rm e}$([\ion{O}{3}]) with respect to SDSS measurements. This large discrepancy might be related to the measurement of [\ion{O}{3}]~\W4363 or due to spatial variations. 
Note that our comparison to SDSS results is due to analysis of the variations of the physical properties in each galaxy with respect to other apertures. Therefore, it does not imply that the temperatures or the different physical conditions of the SDSS sample are the true values since they might also be affected by different bias (i.e., observational problems). 
For the objects with S/N ([\ion{S}{3}]) > 3, we computed $T_{\rm e}$([\ion{S}{3}]) implied by SDSS, LBT,  ESI, and MUSE apertures.  
We find that $T_{\rm e}$([\ion{S}{3}]) calculated using MUSE spectra provides higher temperatures with respect to SDSS results (see Table~\ref{tab:neTe_differences}). 
In particular, J1044+0353 shows a difference of up to 6100 K with respect to $T_{\rm e}$([\ion{S}{3}]) implied by SDSS and LBT spectra. In fact, $T_{\rm e}$([\ion{S}{3}]) derived using LBT also shows a higher value than SDSS. In panel (c) of Fig.~\ref{fig:ne_te}, we show the comparison between $T_{\rm e}$([\ion{S}{3}]) derived using SDSS sample and IFU and long-slit apertures. In that figure, the results using MUSE and LBT for J1044+0353 are represented by a rhomboid and circle, which are joined by a dotted line. We find that the difference between these two measurements and the results of SDSS is up to 6100 K (see Table~\ref{tab:neTe_differences}). These discrepancies might be due to telluric absorption affecting [\ion{S}{3}] \W9069, the only line available in the MUSE spectra. In general, we find differences in $T_{\rm e}$([\ion{S}{3}]) lower than 800 K. 

\begin{figure} 
\begin{center}
    \includegraphics[width=0.45\textwidth, trim=30 0 30 0,  clip=yes]{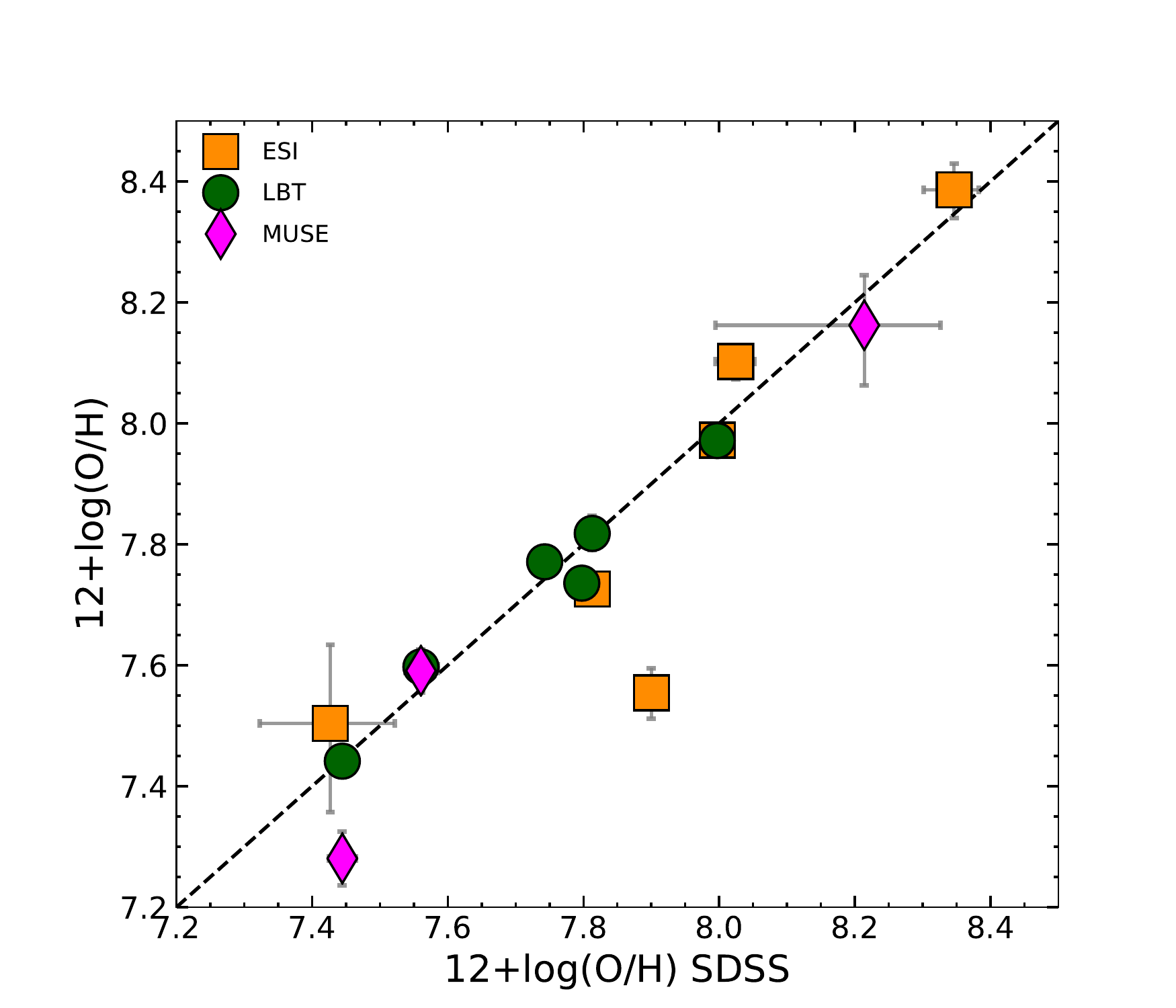}
    \caption{Metallicity comparison between the results implied by SDSS sample and the multiple apertures of IFU and LS (LBT and ESI). The differences in O/H are shown in Table~\ref{tab:metal} for LS and IFU results with respect to SDSS. }
    \label{fig:OH_comparison}
  \end{center}
    \end{figure}

Since metallicity is an important measure in star-forming galaxies, we analyze the impact to use multiple aperture sizes for the same galaxy in the computation of metallicity. One of the main advantages of the sample of CLASSY galaxies studied here is the availability of measurements of electron temperature, which allows a better constraint on the estimated chemical abundance determinations. 
 In Fig.~\ref{fig:OH_comparison}, we show the comparison between the metallicities calculated by the different apertures and those by SDSS aperture. Long-slit spectra correspond to the squares and circles, while the rhomboids correspond to an aperture similar to SDSS and \textit{HST/COS}.  
In general, our results show that LS apertures (most of the sample have aperture sizes of $\approx$ 1\arcsec $\times$ 2\arcsec\,) are consistent with the results implied by the 3.0\arcsec\ aperture of SDSS sample within the uncertainties with differences generally lower than 0.08 dex.

However, we find a  difference in metallicity in using ESI spectra. Such discrepancies are mainly related to the larger differences that we found for  $T_{\rm e}$([\ion{O}{3}]) (see Table~\ref{tab:neTe_differences}).

In principle, it is expected that results obtained from MUSE agree with SDSS given that the apertures map similar emission.  For the three galaxies with MUSE observations, J0021+0052, J1044+0353, and J1148+2546, we find differences in metallicity lower than 0.06 dex, with the exception of J1044+0353 with a difference of 0.16 dex (see Table~\ref{tab:neTe_differences}). The reason for this difference might due to the estimate of $T_{\rm e}$([\ion{O}{3}]) by the temperature relation in Eq. \ref{eq:to3-ts3}. J1044+0353 is an extremely high ionization galaxy showing a large amount of high ionization species \citep{berg21a}. In the total oxygen abundance, the dominant ion is O$^{++}$, implying a strong dependence of the estimated value of $T_{\rm e}$([\ion{O}{3}]) from $T_{\rm e}$([\ion{S}{3}]). Therefore, again another possible explanation for this difference might be associated with observational problems related to the telluric lines affecting the strength of [\ion{S}{3}] \W9069 and/or [\ion{S}{3}] \W9532. After a visual inspection of this sample, we discard such contamination by underlying absorption lines.

In summary, we find a good agreement between IFU and LS apertures for the different physical properties derived for this sample of CLASSY galaxies. The galaxies studied here are mainly dominated by a single bright star formation cluster. Therefore, it is expected that such  differences be minimal. Another point is that  despite instrumental effects and systematic differences between instruments, we are getting the same results. 

On the other hand, since the computation of $T_{\rm e}$ depends on the $T_{\rm e}$-diagnostic available in the emission spectra of each galaxy, it is important to have a reliable constraint of the temperature and ionization structure of the nebular gas. Therefore, this analysis supports comparing the physical properties obtained using optical data with those results used to constrain UV analysis for the aperture of different sizes with similar characteristics to the CLASSY sample \citep{mingozzi22}.

\begin{table*} 
\centering \caption{Electron density, temperature and metallicities differences with respect to SDSS sample. } 
\label{tab:neTe_differences} 
\begin{tabular}{l c c c c c c  } 
\hline 
\hline

\multicolumn{1}{c}{}  & \multicolumn{1}{c}{J0021+0052}   & \multicolumn{1}{c}{ J0808+3948 } & \multicolumn{1}{c}{ J0942+3547 } 
& \multicolumn{1}{c}{ J0944-0038}   & \multicolumn{1}{c}{ J1024+0524} & \multicolumn{1}{c}{ J1044+0353}  \\

\multicolumn{1}{c}{}  & 
\multicolumn{1}{c}{IFU }    &

\multicolumn{1}{c}{ LS}   & 

\multicolumn{1}{c}{ LS }   & 

\multicolumn{1}{c}{ LS}  & 

\multicolumn{1}{c}{ LS}  & 

\multicolumn{1}{c}{IFU | LS}  \\
\hline
\\
$\Delta$$n_{e}$[\ion{S}{2}]  [cm$^{-3}$]   &  $+30$     &   $-100$   &  $-50$ &   $+20$ &  $+20$ &  $-70$ | $-40$    \\
\\

$\Delta$$T_{e}$[\ion{S}{3}] [K]   & N/A  &  N/A   & N/A  &  $-900$ & N/A  & $+6100$ | $+2300$ \\
$\Delta$$T_{e}$[\ion{O}{3}] [K]   & $-200$  & N/A  &  $-200$  &  $-1000$ & N/A &   0  \\
\\

\\
\hline
\multicolumn{1}{c}{}  & \multicolumn{1}{c}{ J1129+2034 }   & 
\multicolumn{1}{c}{ J1132+5722 } & \multicolumn{1}{c}{ J1148+2546 } &
\multicolumn{1}{c}{ J1323-0132}  & \multicolumn{1}{c}{ J1418+2102} & \multicolumn{1}{c}{ J1548+0858 }  \\

\multicolumn{1}{c}{}  & 
\multicolumn{1}{c}{LS}    &

\multicolumn{1}{c}{LS}   & 

\multicolumn{1}{c}{ LS }   & 

\multicolumn{1}{c}{LS}  & \multicolumn{1}{c}{IFU | LS}  &

\multicolumn{1}{c}{LS}    \\
\hline
\\

$\Delta$$n_{e}$[\ion{S}{2}]  [cm$^{-3}$]  &  $+10$ &   $-80$  & $ +10$  &  $-50$ & $-10$ | $+20$ &  $+40$ \\

\\
$\Delta$$T_{e}$[\ion{S}{ 3}] [K]  & $-100$  &  $-700$  & N/A  & N/A  & $+500$ | $-1300$ & N/A \\ 
$\Delta$$T_{\rm e}$[\ion{O}{ 3}] [K]  &  $+200$       &   $-1500$ &  $+100$  &  $+700$ &  N/A |$-100$  &  $-200$  \\    \\

\hline
\end{tabular} 
\tablecomments{Differences in the physical conditions and metallicities, IFU, LS - SDSS. The temperature diagnostics are not available for J0808+3948 for the SDSS spectrum. LS = long-slit.} 
\end{table*} 

\subsection{Aperture analysis using MUSE}
\label{sec:muse-apertures}
As an extra analysis, we have taken advantage of the integrated spectra obtained from MUSE using different aperture sizes for J0021+0052, J1044+0353, and J1418+2102. This new set of spectra comprises sizes between 1.0\arcsec and 7.5\arcsec of diameter with steps of 0.5\arcsec. We analyzed the variations of $E(B-V)$, $n_{\rm e}$, $T_{\rm e}$, ionization parameter, metallicities, SFR, and EWs as result of the different aperture sizes derived from MUSE. 

 We have calculated the SFR using the H$\alpha$ flux measurements obtained in each MUSE apertures.  We use E(B-V) adopted values in each aperture to correct H$\alpha$ fluxes for extinction. 
The luminosity distances were taken from \citet[][see their Table 5]{berg22}, which consider peculiar motions due to the low redshift of the CLASSY galaxies. %
To estimate SFR, we use the expression reported by \citet{kennicutt12}: log SFR($L$(H$\alpha$)) = log $L$(H$\alpha$) -- 41.27. This expression is based on updated stellar models and initial mass function fit by \citet{chabrier03} (mass range 0.1--100 $M_{\odot}$). SFR uncertainties were calculated using error propagation.

\begin{figure*} 
\begin{center}
    \includegraphics[width=0.96\textwidth, trim=30 0 30 0,  clip=yes]{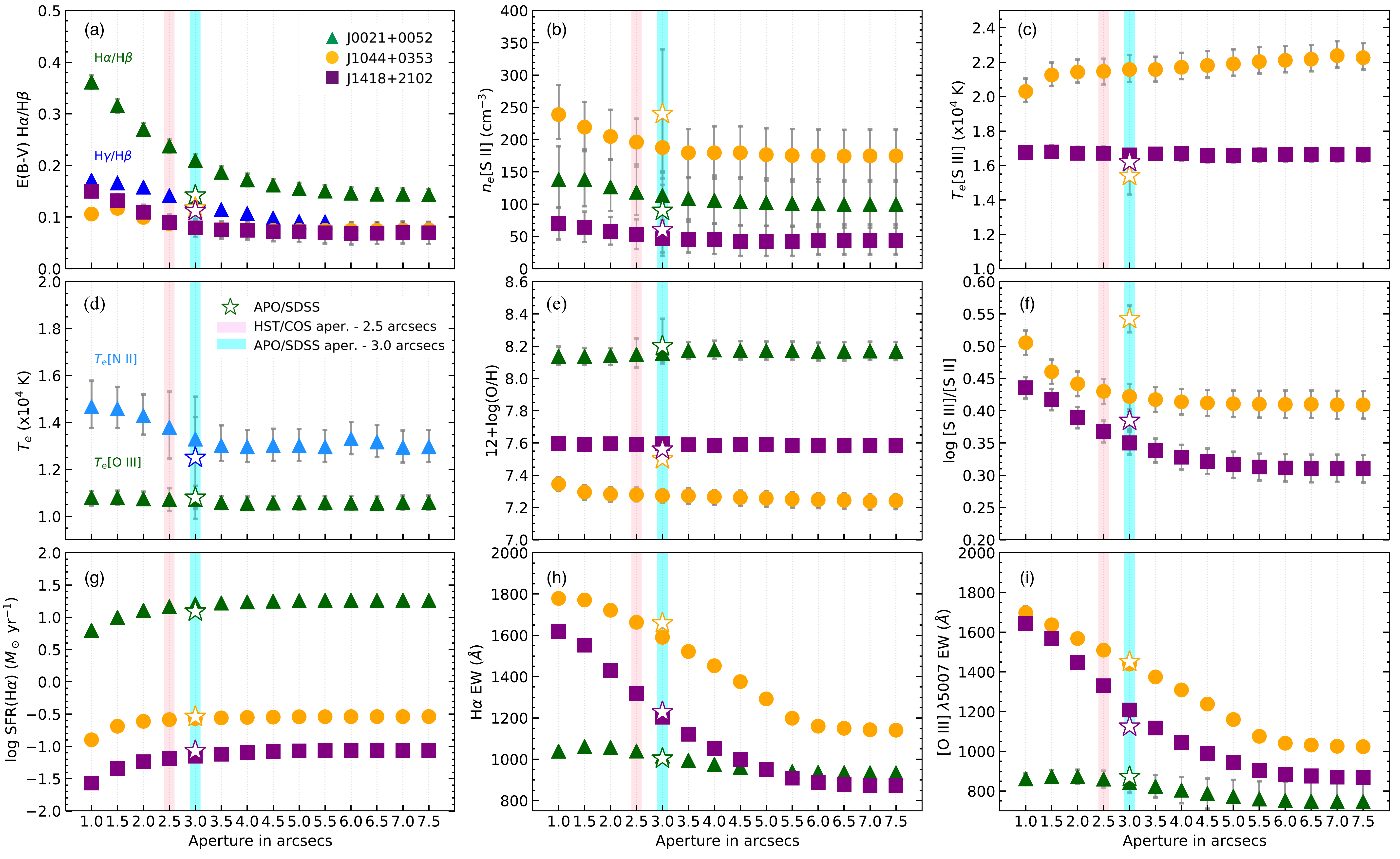}
    \caption{MUSE IFU aperture variations. Integrated MUSE spectra using aperture sizes from 1.0\arcsec\, to 7.5\arcsec\, of diameter in steps of 0.5\arcsec\, for J0021+0052, J1044+0353, and J1418+2102. The different symbols indicate the results for J0021+0052 (triangles), J1044+0353 (circles), and J1418+2102 (squares). For comparison, we add the results using SDSS aperture of 3.0\arcsec\ (stars and cyan bar) spectra for those galaxies. We identify the aperture size of HST/COS of 2.5\arcsec\ (pink bar). Panel (a) shows the reddening derived using H$\alpha$/H$\beta$ and H$\gamma$/H$\beta$ (only for J0021+0052 in blue triangles). Panels (b)-(e) show the variation of electron density ($n_{\rm e}$), electron temperature ($T_{\rm e}$[\ion{O}{3}], $T_{\rm e}$[\ion{N}{2}], and $T_{\rm e}$[\ion{S}{3}]\,) and metallicity. Note that in panels (c) and (d), $T_{\rm e}$[\ion{S}{3}]\ was calculated only for J1044+0353 and J1418+2102, and $T_{\rm e}$[\ion{O}{3}]\, and $T_{\rm e}$[\ion{N}{2}]\ was only determined for J0021+2102, respectively. Panel (f) shows a proxy of the ionization parameter measured as [\ion{S}{3}]~(\W9069+ \W9532)/(\W6717 + \W6731) representative of the low ionization emitting zone \citep[see e.g.,][]{berg21a}. Panels (g)-(i) represent the results of SFR, and the EWs of H$\alpha$ and [\ion{O}{3}], respectively. The results of EW(H$\alpha$) and EW([\ion{O}{3}]) for MUSE and SDSS results for J0021+2102 are multiplied for a factor of 2 for display purposes. Note that J0021+2102 show a smooth variation along the different apertures. The results show that the physical conditions, metallicity, and physical properties converge to constant values at aperture sizes $\sim$ 4.0\arcsec\ of diameter.}
    \label{fig:MUSE_apertures}
  \end{center}
    \end{figure*}

In Fig.~\ref{fig:MUSE_apertures}, we show the results for the integrated MUSE spectra using a different aperture size. The triangles, circles, and squares represent J0021+0052, J1044+0353, and J1418+2102, respectively. For comparison, we have added the results derived from SDSS spectra, which are shown in each panel of Fig.~\ref{fig:MUSE_apertures} with empty stars with the same color as their respective galaxy. The two vertical bars represent the aperture size of HST/COS and SDSS of 2.5\arcsec\ and 3.0\arcsec\, respectively. 
Overall, Fig.~\ref{fig:MUSE_apertures} shows that the variations of the galaxy properties convergent at an aperture size of ~ 4\arcsec , implying that most of the flux can be in-closed at similar aperture sizes to SDSS (3\arcsec ) and COS (2.5\arcsec ) apertures for these galaxies.
In panels (a) to (e), we compared each individual spectrum as a function of the extinction, the physical conditions ($n_{\rm e}$ and $T_{\rm e}$) and metallicity. In general, we find consistent results through the multiple aperture sizes. We note that for J0021+0052 the extinction calculated using H$\alpha$/H$\beta$ shows a difference of 0.15 dex when 1.0\arcsec\ and 3.0\arcsec\ aperture are compared, increasing such difference as the aperture becomes bigger. However, we find a smooth variation of $E(B-V)$ for J10144+0353 and J1418+2102 with respect to the different aperture sizes. We also calculated $E(B-V)$ using H$\gamma$/H$\beta$ for J0021+0052 (see blue triangles in panel (a) of Fig.~\ref{fig:MUSE_apertures}.), showing a similar behavior in comparison to the other two galaxies in this analysis. In fact,  the results for SDSS show a good agreement with those obtained using 2.5\arcsec aperture size. 

We also find that the electron density and temperature, and metallicity show slight variations in their results for the different aperture sizes (see panels (b)-(e) of Fig.~\ref{fig:MUSE_apertures}). For J1044+0353 and J1418+2102, the only temperature diagnostic available is $T_{\rm e}$([\ion{S}{3}]), whose results are in excellent agreement with the rest of MUSE aperture sizes implied in this analysis.
For J1044+0353, the result for $T_{\rm e}$([\ion{S}{3}] show a large difference between MUSE and SDSS data (see Table~\ref{tab:neTe_differences}). In Sec.~\ref{sec:den_ten_metal}, we stress that such a difference could be due attributed to the telluric features affecting the [\ion{S}{3}]\, \W9069 emission line.

In Panel (d) of Fig.~\ref{fig:MUSE_apertures}, we show the results for $T_{\rm e}$([\ion{O}{3}]) and $T_{\rm e}$([\ion{N}{2}]) for J0021+0052. The results for $T_{\rm e}$([\ion{O}{3}]) are consistent between the different apertures. However, the variation of $T_{\rm e}$([\ion{N}{2}]) is slightly larger increasing to small aperture sizes. Interestingly, we have calculated high values of  $T_{\rm e}$([\ion{N}{2}]) with respect to those obtained for $T_{\rm e}$([\ion{O}{3}]). On the other hand, panel (e) shows small variations of metallicity as the aperture size increases. Moreover, SDSS spectra show a good agreement with the values of O/H derived either using an aperture of 2.5\arcsec or 3.0\arcsec. 
In panel (f) of Fig.~\ref{fig:MUSE_apertures}, we compare the variations of the ionization parameter measured as a proxy of S3S2 = [\ion{S}{3}]~(\W9069+ \W9532)/[\ion{S}{2}]~(\W6717 + \W6731) \citep[see e.g.,][]{berg21a, mingozzi22}, which trace the low ionization emitting zone of the nebula. We have used this approach  because [\ion{S}{2}] and [\ion{S}3] are directly available for J1044+0353 and J1418+2102 in MUSE spectra. We find that log(S3S2) decreases until reach a convergence when the aperture size is also $\sim$ 4\arcsec of diameter. 
For J1418+0858, we find a small difference between the MUSE and SDSS results of 0.03 dex for an aperture size of 3
\arcsec. However, the log(S3S2) obtained with MUSE shows a large difference with respect to the SDSS value (see also panel (c) in Fig.~\ref{fig:MUSE_apertures}), probably associanted with the measurement of the [\ion{S}{3}]~(\W9069) line as we have discussed in Sec.~\ref{sec:den_ten_metal}.  

Finally, we compare SFR and EWs of H$\alpha$ and [\ion{O}{3}]\, for each aperture, see panels (f) to (h) of Fig.~\ref{fig:MUSE_apertures}. Although, we find small changes in the inferred SFRs at a given aperture size, note that for apertures smaller than 1.5 we obtained slightly lower values of SFR for J0021+0052, J1044+0353, and J1418+2102. Moreover, the SFR and EWs converge when the aperture size is $\geq$ 5.0 \arcsec .
We find high values of EWs for H$\alpha$ and [\ion{O}{3}]\ for those galaxies, with an excellent agreement with those results derived from SDSS spectra. We also note that EWs of  H$\alpha$ and [\ion{O}{3}]\ of J0021+0052 shows a drop for spectra with an aperture size of 1\arcsec (due to the scale of panels (g) and (H) of Fig.~\ref{fig:MUSE_apertures} such a structure is not visible) followed by an increase of EW at an aperture of 1.5\arcsec. Such behavior might be related to the complex structure into the core of J0021+0052. 

\section{Summary and Conclusions} 
\label{sec:conclusion}
A key goal of the CLASSY sample is to provide a unified picture of the
stars and gas with in nearby star-forming galaxies, which is largely
derived from their FUV spectra.
However, optical spectra that are well matched to the FUV spectra 
are also needed to derive a number of important nebular properties.  
We, therefore, analyzed the impact of aperture differences on the determination 
of nebular properties of 12 local star-forming galaxies at 
$z < 0.098$ for the CLASSY sample.
This sample was chosen to have multiple optical spectra, including 
circular aperture, long-slit, and IFU spectroscopy. 
Specifically, we investigated the nebular properties of 
reddening, electron density and temperature, metallicities, ionization, and nebular to
stellar emission as probed by H$\alpha$ and [\ion{O}{3}] \W5007 equivalent width. 
In principle, the 3\arcsec diameter SDSS aperture is expected to cover most 
of the emission of CLASSY galaxies and is well-matched to the FUV spectra 
observed with the HST/COS 2\farcs5 aperture. 
Therefore, we use the SDSS spectra as our base of comparison for our the other optical spectra.
Additionally, the IFU spectra allow us to inspect how the derived properties change
as a function of aperture-extraction size, while eliminating any instrument effects. 
We summarize our main conclusions as follows. 

1.~We have calculated the reddening using three different Balmer ratios 
(H$\alpha$/H$\beta$, H$\gamma$/H$\beta$, and H$\delta$/H$\beta$) and found
that the error-weighted average E(B--V) value was insensitive to aperture size
of different spectra for our sample, with a median difference of less than 0.1 dex.

However, using the IFU observations of 3 CLASSY galaxies, we find that the E(B-V) values derived from individual Balmer line ratios decrease (by up to 53\%) with increasing aperture size, with the most significant change occurring in the center of the galaxies.

2.~We calculated electron densities [\ion{S}{2}] \W\W6717,6731 and temperatures from 
multiple auroral lines and found them to be insensitive to aperture size for our sample.
In particular, investigating a range of aperture-extraction sizes from the 
IFU observations, we find that values change most significantly in the center 
of the galaxies, and level out near the COS aperture radius of 2\farcs5.
Similar results were found for reddening and metallicity, 

3.~We find a good agreement between the metallicities derived using the various
aperture spectra for the same galaxy, with differences of $<$0.1 dex. 
Such small differences imply that the metallicity calculated from the optical spectra
is representative of the region sampled in the FUV with COS for our CLASSY sample 
of the star-forming galaxies.

To summarize, we find that the aperture effects on inferred nebular properties
are minimal for the CLASSY sample of star-forming galaxies.
Here we want to stress that despite the specific instrumental effects imprinted 
on the spectra, we find quite similar results. 
These results demonstrate the appropriateness of comparing the physical 
properties obtained in the optical for compact, highly-star-forming galaxies
with those using the 2.5\arcsec\ aperture of HST/COS \citep[][]{mingozzi22}.

\clearpage
The CLASSY team thanks the referee for thoughtful feedback that improved the paper.
KZA-C and DAB are grateful for the support for this program, HST-GO-15840, that was provided by 
NASA through a grant from the Space Telescope Science Institute, which is operated by the Associations of Universities for Research in Astronomy, 
Incorporated, under NASA contract NAS5-26555. 
The CLASSY collaboration extends special gratitude to the Lorentz Center for useful discussions 
during the "Characterizing Galaxies with Spectroscopy with a view for JWST" 2017 workshop that led 
to the formation of the CLASSY collaboration and survey.
The CLASSY collaboration thanks the COS team for all their assistance and advice in the 
reduction of the COS data.
KZA-C also thanks Noah Rogers for his help with the MODS data reduction for LBT spectra and V\'ictor Pati\~no-\'Alvarez and H\'ector Ibarral-Medel for their helpful discussions in the use of {\sc Starlight} code.
BLJ is thankful for support from the European Space Agency (ESA).
JB acknowledges support by Fundação para a Ciência e a Tecnologia (FCT) through the research grants UIDB/04434/2020 and UIDP/04434/2020, through work contract No. 2020.03379.CEECIND, and through FCT project PTDC/FISAST/4862/2020. RA acknowledges support from ANID Fondecyt Regular 1202007. 

This work also uses observations obtained with the Large 11 Binocular Telescope (LBT). The LBT is an international collaboration among institutions in the United States, Italy and Germany. LBT Corporation partners are: The University of Arizona on behalf of the Arizona Board of Regents; Istituto Nazionale di Astrofisica, Italy; LBT Beteiligungsge-sellschaft, Germany, representing the Max-Planck Society,The Leibniz Institute for Astrophysics Potsdam, and Heidelberg University; The Ohio State University, University of
19 Notre Dame, University of Minnesota, and University of Virginia. 

The STARLIGHT project is supported by the Brazilian agencies CNPq, CAPES and FAPESP and by the France-Brazil CAPES/Cofecub program.
Funding for SDSS-III has been provided by the Alfred P. Sloan Foundation, the Participating Institutions, the National Science Foundation, and the U.S. Department of Energy Office of Science. 

The SDSS-III web site is http://www.sdss3.org/.
SDSS-III is managed by the Astrophysical Research Consortium for the Participating Institutions of the SDSS-III Collaboration including the University of Arizona, the Brazilian Participation Group, Brookhaven National Laboratory, Carnegie Mellon University, University of Florida, the French Participation Group, the German Participation Group, Harvard University, the Instituto de Astrofisica de Canarias, the Michigan State/Notre Dame/JINA Participation Group, Johns Hopkins University, Lawrence Berkeley National Laboratory, Max Planck Institute for Astrophysics, Max Planck Institute for Extraterrestrial Physics, New Mexico State University, New York University, Ohio State University, Pennsylvania State University, University of Portsmouth, Princeton University, the Spanish Participation Group, University of Tokyo, University of Utah, Vanderbilt University, University of Virginia, University of Washington, and Yale University.

This work also uses the services of the ESO Science Archive Facility,
observations collected at the European Southern Observatory under 
ESO programmes 096.B-0690, 0103.B-0531, 0103.D-0705, and 0104.D-0503. 

\bibliography{mybib}
\clearpage
\appendix
\section{Emission-Line Intensities}
Here we include the following tables:

\begin{itemize}
    \item Table~\ref{tab:intensitiesSDSS}: Emission line intensities measured for the APO/SDSS sample of 12 star-forming galaxies. 
    \item Table~\ref{tab:intensities_ESI}: Emission line intensities measured for the Kekc/ESI sample of six star-forming galaxies. 
    \item Table~\ref{tab:intensities_MUSE}: Emission line intensities measured for the VLT/MUSE sample of three star-forming galaxies. 
\end{itemize}

The optical emission lines for the entire CLASSY sample are reported in \citet{mingozzi22}.  

\begin{deluxetable*}{c c c c c c c c c c c c c }
\tablewidth{0pt}
\caption{Deredened emission line intensities for the APO/SDSS spectra for 12 CLASSY galaxies.
\label{tab:intensitiesSDSS}}
\tablehead{
Wavelength & Ion & J0021+0052 & J0808+3948 & J0942+3547 & J0944-0038 & J1024+0542 & J1044+0352\\
(\AA)      &                  &            &           &            &            &  
}
\startdata
3726.04             &    [\ion{O}{2}]     &   $87.0\pm 3.1$       &  $103.49\pm9.0$         &   \nodata            &      \nodata             &  50.8 $\pm$ 1.6      &      \nodata           \\
3728.80             &    [\ion{O}{2}]     &   $102.6\pm3.2$       &  \nodata                &   \nodata            &      \nodata             &  65.8 $\pm$ 1.7      &      \nodata           \\
4101.71             &    H$\delta$        &   $26.29 \pm 0.51$    &  $27.67 \pm3.14$        &   27.35 $\pm$ 0.41   &    26.45 $\pm$ 0.33      &  25.41 $\pm$ 0.35    &      27.82 $\pm$ 0.37  \\
4340.44             &    H$\gamma$        &   $47.33 \pm 0.57$    &  $54.71 \pm3.7 $        &   48.06 $\pm$ 0.44   &    47.13 $\pm$ 0.42      &  45.42 $\pm$ 0.36    &      47.8 $\pm$ 0.46   \\
4363.21             &    [\ion{O}{3}]     &   $3.76  \pm 0.48$    &  \nodata                &   6.8 $\pm$ 0.29     &    11.86 $\pm$ 0.24      &  9.15 $\pm$ 0.29     &       13.7 $\pm$ 0.29   \\
4861.35             &    H$\beta$         &   $100.0 \pm 1.0$     &  100.0 $\pm$ 5.0        &   100.0 $\pm$ 1.0    &    100.0 $\pm$ 1.0       &  100.0 $\pm$ 0.76    &      100.0 $\pm$ 1.0   \\
4958.61             &    [\ion{O}{3}]     &   $151.23 \pm 2.25$   &  18.09 $\pm$ 2.19       &   169.03 $\pm$ 1.75  &    185.89 $\pm$ 2.33     &  173.17 $\pm$ 1.73   &     147.66 $\pm$ 2.79 \\
5006.84             &    [\ion{O}{3}]     &   $455.26 \pm 4.05$   &  55.33 $\pm$ 2.91       &   497.56 $\pm$ 3.94  &    555.47 $\pm$ 3.96     &  520.04 $\pm$ 3.33   &     \nodata   \\
5754.64             &    [\ion{N}{2}]     &   $0.57   \pm 0.24$   &  \nodata                &   0.48 $\pm$ 0.21    &    \nodata               &  \nodata             &                \nodata           \\
6312.06             &    [\ion{S}{3}]     &   $1.04   \pm 0.15$   &  \nodata                &   1.77 $\pm$ 0.13    &    1.46 $\pm$ 0.04       &  1.72 $\pm$ 0.1      &        0.69 $\pm$ 0.07   \\
6562.79             &    H$\alpha$        &   $287.2  \pm 3.76$   &  296.0 $\pm$ 16.1       &   281.53 $\pm$ 3.05  &    \nodata               &  273.66 $\pm$ 2.39   &      288.6 $\pm$ 4.0   \\
6583.45             &    [\ion{N}{2}]     &   $23.52  \pm 2.01$   &  207.7 $\pm$ 11.84      &   10.96 $\pm$ 0.89   &    3.37 $\pm$ 4.57       &  5.96 $\pm$ 1.1      &        0.93 $\pm$ 2.49   \\
6716.44             &    [\ion{S}{2}]     &   $16.68  \pm 0.37$   &  19.18 $\pm$ 1.4        &   16.43 $\pm$ 0.24   &    7.39 $\pm$ 0.1        &  11.44 $\pm$ 0.16    &      2.42 $\pm$ 0.08   \\
6730.82             &    [\ion{S}{2}]     &   $12.64  \pm 0.35$   &  24.61 $\pm$ 1.64       &   11.91 $\pm$ 0.21   &    5.74 $\pm$ 0.09       &  8.56 $\pm$ 0.15     &       2.02 $\pm$ 0.08   \\
7319.92             &    [\ion{O}{2}]     &   $2.22   \pm 0.14$   &  \nodata                &   1.63 $\pm$ 0.13    &    1.61 $\pm$ 0.04       &  1.69 $\pm$ 0.11     &       0.53 $\pm$ 0.06   \\
7339.79             &    [\ion{O}{2}]     &   $1.98   \pm 0.14$   &  \nodata                &   1.46 $\pm$ 0.13    &    1.29 $\pm$ 0.04       &  1.42 $\pm$ 0.11     &       0.49 $\pm$ 0.06   \\
9068.60             &    [\ion{S}{3}]     &   \nodata            &  \nodata                 &   \nodata            &    10.94 $\pm$ 0.32      &  \nodata            &        4.47 $\pm$ 0.23   \\
\hline
$E(B-V)$            &   & 0.135$\pm$0.009  & 0.296$\pm$0.036  & 0.021$\pm$0.006 & 0.193$\pm$0.009 & 0.025$\pm$0.006 & 0.072$\pm$0.008   \\
$ F_{H\beta}$        &   &  292.7$\pm$2.2 &   47.7$\pm$1.7 &  214.5$\pm$1.6 &  813.5$\pm$4.7 &  331.9$\pm$1.8 &  471.9$\pm$3.4  \\
\hline
Wavelength & Ion & J1129+2034 & J1132+5722 & J1148+2546   & J1323-0132 &  J1418+2546 & J1545+0858\\
(\AA)      &                  &            &           &            &            & \\
\hline
3726.04             &    [\ion{O}{2}]    &  \nodata     &          \nodata     &         59.4 $\pm$ 1.4  &      \nodata      &            \nodata     &          35.5 $\pm$ 1.5    \\
3728.80             &    [\ion{O}{2}]    &  \nodata     &          \nodata     &         78.9 $\pm$ 1.5  &      \nodata      &            \nodata     &          43.1 $\pm$ 1.5    \\
4101.71             &    H$\delta$       &  \nodata     &          26.98 $\pm$ 0.81  &   26.26 $\pm$ 0.34  &       24.51 $\pm$ 0.45  &       \nodata    &           26.79 $\pm$ 0.31   \\
4340.44             &    H$\gamma$       &  48.55 $\pm$ 0.45   &   46.64 $\pm$ 0.61  &   46.52 $\pm$ 0.36  &       45.97 $\pm$ 0.45  &       46.5 $\pm$ 0.41   &    46.09 $\pm$ 0.38   \\
4363.21             &    [\ion{O}{3}]    &  3.4 $\pm$ 0.21     &   5.36 $\pm$ 0.72   &   9.91 $\pm$ 0.21   &       19.17 $\pm$ 0.35  &       12.78 $\pm$ 0.39  &    12.49 $\pm$ 0.34   \\
4861.35             &    H$\beta$        &  100.0 $\pm$ 1.07   &   100.0 $\pm$ 1.05  &   100.0 $\pm$ 0.91  &       100.0 $\pm$ 1.15  &       100.0 $\pm$ 0.61  &    100.0 $\pm$ 0.64   \\
4958.61             &    [\ion{O}{3}]    &  156.54 $\pm$ 3.88  &   69.7 $\pm$ 1.02  &    200.24 $\pm$ 2.66  &      250.37 $\pm$ 5.39  &      155.82 $\pm$ 1.33  &   183.87 $\pm$ 1.35  \\
5006.84             &    [\ion{O}{3}]    &  \nodata   &   202.24 $\pm$ 1.82  &  607.25 $\pm$ 4.76  &      761.02 $\pm$ 8.07  &      452.0 $\pm$ 2.5  &     543.64 $\pm$ 2.84  \\
6312.06             &    [\ion{S}{3}]    &  1.6 $\pm$ 0.05     &   0.99 $\pm$ 0.23  &    1.32 $\pm$ 0.09  &        \nodata  &                1.02 $\pm$ 0.07  &     1.22 $\pm$ 0.06    \\
6562.79             &    H$\alpha$       &  293.49 $\pm$ 3.97  &   276.18 $\pm$ 3.3  &   278.38 $\pm$ 2.59  &      269.53 $\pm$ 3.05  &      275.16 $\pm$ 2.13  &   277.05 $\pm$ 1.86  \\
6583.45             &    [\ion{N}{2}]    &  11.61 $\pm$ 2.11   &   4.13 $\pm$ 1.4  &     6.26 $\pm$ 0.88  &        1.01 $\pm$ 0.86  &        2.04 $\pm$ 0.96  &     3.2 $\pm$ 0.62     \\
6716.44             &    [\ion{S}{2}]    &  14.9 $\pm$ 0.27    &   12.74 $\pm$ 0.3  &    11.79 $\pm$ 0.17  &       2.11 $\pm$ 0.17  &        5.02 $\pm$ 0.09  &     6.19 $\pm$ 0.08    \\
6730.82             &    [\ion{S}{2}]    &  11.2 $\pm$ 0.24    &   9.94 $\pm$ 0.29  &    8.98 $\pm$ 0.15  &        2.2 $\pm$ 0.17  &         3.66 $\pm$ 0.08  &     4.67 $\pm$ 0.07    \\
7319.92             &    [\ion{O}{2}]    &  2.11 $\pm$ 0.06    &   2.38 $\pm$ 0.26  &    1.76 $\pm$ 0.09  &        0.43 $\pm$ 0.07  &        0.95 $\pm$ 0.06  &     0.94 $\pm$ 0.06    \\
7339.79             &    [\ion{O}{2}]    &  1.7 $\pm$ 0.06     &   2.19 $\pm$ 0.25  &    1.54 $\pm$ 0.08  &        0.4 $\pm$ 0.05  &         0.67 $\pm$ 0.06  &     0.85 $\pm$ 0.06    \\
9068.60             &    [\ion{S}{3}]    &  21.36 $\pm$ 0.6    &   6.73 $\pm$ 0.28  &    \nodata &                 \nodata &                 6.06 $\pm$ 0.23  &     \nodata           \\
\hline
$E(B-V)$            &   & 0.191$\pm$0.009  & 0.038$\pm$0.008 & 0.119$\pm$0.006 & 0.147$\pm$0.007 & 0.116$\pm$0.005 & 0.156$\pm$0.004  \\
$F_{H\beta}$        &   &  386.3$\pm$2.9 &  86.5$\pm$0.6  &  380.0$\pm$2.4 &  168.5$\pm$1.4 &  270.0$\pm$1.2 &  555.3$\pm$ 2.5    \\
\enddata
\tablecomments{Intensity ratios are reported with respect to $I$(H$\beta$) = 100, where the observed
H$\beta$ fluxes (in units of $10^{-16}$ erg s$^{-1}$ cm$^{-2}$) are reported in the second to last row.
The color excess value, $E(B-V)$, used to reddening correct the line ratios
are listed in the last row. }
\end{deluxetable*}
\begin{deluxetable*}{c c c c c c c c }
\tablewidth{0pt}
\caption{Deredened emission line intensities measured for the Keck/ESI spectra for five CLASSY galaxies.}
\label{tab:intensities_ESI}
\tablehead{
Wavelength & Ion & J0942+3547  & J0944-0038  & J1024+0524  & J1129+2034  & J1132+5722  & J1148+2546\\
(\AA) &  &   &  &   &   &   &
}
\startdata
4101.71             &    H$\delta$  &  27.96 $\pm$ 0.29        &    28.65 $\pm$ 0.16     &   27.33 $\pm$ 0.37    &  24.92 $\pm$ 0.46    &  27.24 $\pm$ 1.26  &   28.83 $\pm$ 0.2    \\
4340.44             &    H$\gamma$  &  50.6 $\pm$ 0.4          &    49.13 $\pm$ 0.21     &  \nodata       &  47.94 $\pm$ 0.57    &  45.02 $\pm$ 1.18  &   0.0 $\pm$ 0.0      \\
4363.21             &    [\ion{O}{3}]   &  6.22 $\pm$ 0.24         &    13.14 $\pm$ 0.13     &   13.82 $\pm$ 0.38    &  3.41 $\pm$ 0.23     &  5.11 $\pm$ 1.0    &   10.62 $\pm$ 0.22   \\
4861.35             &    H$\beta$   &  100.0 $\pm$ 0.64        &    100.0 $\pm$ 0.32     &   100.0 $\pm$ 1.09    &  100.0 $\pm$ 0.59    &  100.0 $\pm$ 1.5   &   100.0 $\pm$ 0.49   \\
4958.61             &    [\ion{O}{3}]   &  166.42 $\pm$ 1.65       &    190.59 $\pm$ 1.45    &   182.73 $\pm$ 2.4    &  158.29 $\pm$ 6.23   &  77.39 $\pm$ 1.53  &   215.2 $\pm$ 1.36   \\
5754.64             &    [\ion{N}{2}]    &  0.17 $\pm$ 0.05         &    0.07 $\pm$ 0.02      &   0.14 $\pm$ 0.04     &  0.21 $\pm$ 0.03     &  0.26 $\pm$ 0.14   &   0.1 $\pm$ 0.03     \\
6312.06             &    [\ion{S}{3}]   &  1.82 $\pm$ 0.08         &    1.52 $\pm$ 0.02      &   1.51 $\pm$ 0.05     &  1.86 $\pm$ 0.04     &  0.88 $\pm$ 0.15   &   1.27 $\pm$ 0.03    \\
6562.79             &    H$\alpha$  &  282.22 $\pm$ 2.3        &    287.67 $\pm$ 1.18    &   281.83 $\pm$ 3.24   &  \nodata      &  277.04 $\pm$ 4.19 &   279.95 $\pm$ 1.6   \\
6583.45             &    [\ion{N}{2}]    &  11.23 $\pm$ 0.99        &    3.44 $\pm$ 0.63      &   5.55 $\pm$ 1.12     &  13.91 $\pm$ 0.56    &  3.74 $\pm$ 0.55   &   5.42 $\pm$ 0.57    \\
6716.44             &    [\ion{S}{2}]    &  17.77 $\pm$ 0.21        &    8.12 $\pm$ 0.05      &   11.48 $\pm$ 0.37    &  18.43 $\pm$ 0.29    &  12.82 $\pm$ 0.26  &   11.12 $\pm$ 0.1    \\
6730.82             &    [\ion{S}{2}]    &  12.49 $\pm$ 0.19        &    6.18 $\pm$ 0.05      &   8.2 $\pm$ 0.36      &  13.96 $\pm$ 0.23    &  9.32 $\pm$ 0.21   &   8.65 $\pm$ 0.09    \\
7319.92             &    [\ion{O}{2}]    &  2.16 $\pm$ 0.15         &    1.69 $\pm$ 0.02      &   1.71 $\pm$ 0.34     &  2.63 $\pm$ 0.06     &  1.91 $\pm$ 0.15   &   1.48 $\pm$ 0.08    \\
7339.79             &    [\ion{O}{2}]    &  1.75 $\pm$ 0.15         &    1.3 $\pm$ 0.02       &   1.36 $\pm$ 0.34     &  2.06 $\pm$ 0.05     &  1.5 $\pm$ 0.14    &   1.39 $\pm$ 0.08    \\
9068.60             &    [\ion{S}{3}]   &  18.81 $\pm$ 0.27        &    10.24 $\pm$ 0.12     &   7.47 $\pm$ 0.2      &  29.75 $\pm$ 0.81    &  5.53 $\pm$ 0.16   &   9.52 $\pm$ 0.12    \\
\hline
$E(B-V)$            &   & $0.036\pm0.008$   & $0.134\pm0.003$     & $0.055\pm0.008$   & $0.125\pm0.013$ & $0.054\pm0.010$ & $0.071\pm0.004$   \\
$F_{H\beta}$        &   & $195.8 \pm1.4$ & $1073.6\pm2.4 $ & $309.5 \pm2.4$ & $552.0\pm2.3 $ & $686.7\pm7.3$ & $357.8\pm1.2$ \\
\enddata
\tablecomments{Intensity ratios are reported with respect to $I$(H$\beta$) = 100, where the observed
H$\beta$ fluxes (in units of 10$^{-16}$ erg s$^{-1}$ cm$^{-2}$) are reported in the second to last row.
The color excess value, $E(B-V)$, used to reddening correct the line ratios
are listed in the last row. }
\end{deluxetable*}
\begin{deluxetable}{c c c c c c c c }
\tablewidth{0pt}
\caption{Deredened emission line intensities measured for the VLT/MUSE spectra for Three CLASSY galaxies.
\label{tab:intensities_MUSE}}
\tablehead{
Wavelength & Ion & J021+0052  & J1044+0353 & J1418+2102\\
(\AA)      &    &             &            &             }
\startdata
4340.44             &    H$\gamma$    &    46.36 $\pm$ 0.64   &   \nodata                 &  \nodata                 \\
4363.21             &    [\ion{O}{3}]     &    3.44 $\pm$ 0.47    &   \nodata&  \nodata                \\
4861.35             &    H$\beta$     &    100.0 $\pm$ 1.13   &   100.0 $\pm$ 1.0    &   100.0 $\pm$ 1.0   \\
4958.61             &    [\ion{O}{3}]     &    146.75 $\pm$ 2.27  &   147.43 $\pm$ 1.16   &   160.58 $\pm$ 1.86  \\
5006.84             &    [\ion{O}{3}]    &    440.14 $\pm$ 4.21  &   436.24 $\pm$ 2.39   &   470.51 $\pm$ 3.08  \\
5754.64             &    [\ion{N}{2}]      &    0.67 $\pm$ 0.11    &   \nodata      &   \nodata                 \\
6312.06             &    [\ion{S}{3}]     &    1.25 $\pm$ 0.09    &   0.76 $\pm$ 0.01     &   1.01 $\pm$ 0.02    \\
6562.79             &    H$\alpha$    &     283.54 $\pm$ 4.69  &   273.4 $\pm$ 2.5     &   277.26 $\pm$ 2.29  \\
6583.45             &    [\ion{N}{2}]      &    21.74 $\pm$ 2.21   &   0.91 $\pm$ 0.091      &   1.91 $\pm$ 0.69    \\
6716.44             &    [\ion{S}{2}]      &    16.2 $\pm$ 0.37    &   2.34 $\pm$ 0.04     &   5.12 $\pm$ 0.06    \\
6730.82             &    [\ion{S}{2}]      &    12.56 $\pm$ 0.34   &   1.9 $\pm$ 0.04      &   3.74 $\pm$ 0.06    \\
7319.92             &    [\ion{O}{2}]      &    2.05 $\pm$ 0.07    &   0.59 $\pm$ 0.01     &   0.9 $\pm$ 0.02     \\
7339.79             &    [\ion{O}{2}]      &    1.72 $\pm$ 0.06    &   0.49 $\pm$ 0.01     &   0.79 $\pm$ 0.02    \\
9068.60             &    [\ion{S}{3}]     &   \nodata                 &   3.15 $\pm$ 0.16     &   5.78 $\pm$ 0.2     \\
\hline
\hline
$E(B-V)$            &   &  $0.214\pm0.011$ &   $0.088\pm0.007$  & $0.091\pm0.006$ \\
$F_{H\beta}$     &   & $187.0\pm1.5$  &  $404.6\pm2.0$  & $211.0\pm1.1$    \\
\enddata
\tablecomments{Intensity ratios are reported with respect to $I$(H$\beta$) = 100, where the observed
H$\beta$ fluxes (in units of 10$^{-16}$ erg s$^{-1}$ cm$^{-2}$) are reported in the second to last row.
The color excess value, $E(B-V)$, used to reddening correct the line ratios
are listed in the last row. }
\end{deluxetable}

\end{document}